\documentclass[prd,nofootinbib,onecolumn,preprint,11pt]{revtex4}

\usepackage{graphicx}
\usepackage{bm}
\usepackage{float}
\usepackage{subfigure}
\usepackage{amsmath, amssymb}
\usepackage{amsfonts}
\usepackage{color}
\usepackage{slashed}
\usepackage{fancyhdr}

\newcommand{\mpl}{M_{\rm pl}}
\newcommand{\GN}{G_{\rm N}}
\newcommand{\gb}{\bar{g}}

\newcommand{\sbar}{\bar{\sigma}}

\newcommand{\dd}{\text{d}}

\begin{document}

\title{Transverse-Traceless Gravitational Waves In A Spatially Flat FLRW Universe: Causal Structure from Dimension Reduction}

\author{Yi-Zen Chu}
\affiliation{
Department of Physics, University of Minnesota, 1023 University Dr., Duluth, MN 55812, USA
}

\begin{abstract}
\noindent This work was mainly driven by the desire to explore, to what extent embedding some given geometry in a higher dimensional flat one is useful for understanding the causal structure of classical fields traveling in the former, in terms of that in the latter. We point out, in the 4-dimensional (4D) spatially flat Friedmann-Lema\^{i}tre-Robertson-Walker universe, that the causal structure of transverse-traceless (TT) gravitational waves can be elucidated by first reducing the problem to a 2D Minkowski wave equation with a time dependent potential, where the relevant Green's function is pure tail -- waves produced by a physical source propagate strictly within the null cone. By viewing this 2D world as embedded in a 4D one, the 2D Green's function can also be seen to be sourced by a cylindrically symmetric scalar field in 3D. From both the 2D wave equation as well as the 3D scalar perspective, we recover the exact solution of the 4D graviton tail, for the case where the scale factor written in conformal time is a power law. There are no TT gravitational wave tails when the universe is radiation dominated because the background Ricci scalar is zero. In a matter dominated one, we estimate the amplitude of the tail to be suppressed relative to its null counterpart by both the ratio of the duration of the (isolated) source to the age of the universe $\eta_0$, and the ratio of the observer-source spatial distance (at the observer's time) to the same $\eta_0$. In a universe driven primarily by a cosmological constant, the tail contribution to the background geometry $a[\eta]^2 \eta_{\mu\nu}$ after the source has ceased, is the conformal factor $a^2$ times a spacetime-constant symmetric matrix proportional to the spacetime volume integral of the TT part of the source's stress-energy-momentum tensor. In other words, massless spin-2 gravitational waves exhibit a tail-induced memory effect in 4D de Sitter spacetime.
\end{abstract}

\maketitle

\section{Motivation and Introduction}

The geometry of our universe appears to be well described by Einstein's equations with a cosmological constant $\Lambda = 3H^2$,
\begin{align}
\label{EinsteinEquatin}
G^\mu_{\phantom{\mu}\nu} - \Lambda \delta^\mu_{\phantom{\mu}\nu} = 8 \pi \GN T^\mu_{\phantom{\mu}\nu} .
\end{align}
At zeroth order and at very large scales, $T^\mu_{\phantom{\mu}\nu} = \overline{T}^\mu_{\phantom{\mu}\nu}$ contains an isotropic and homogeneous background matter-energy distribution that drives the evolution of a 4-dimensional (4D) spatially flat Friedmann-Lema\^{i}tre-Robertson-Walker (FLRW) geometry parametrized by conformal time $\eta$ and $3$ spatial coordinates $\vec{x}$, i.e.,
\begin{align}
\label{4DFlatFLRW}
\gb_{\mu\nu}[\eta] \equiv a[\eta]^2 \eta_{\mu\nu}, \qquad
\text{diag}[1,-1,-1,-1] .
\end{align}
At first order, $T^\mu_{\phantom{\mu}\nu} = \overline{T}^\mu_{\phantom{\mu}\nu} + \delta T^\mu_{\phantom{\mu}\nu}$ also includes the inhomogeneous matter perturbations $\delta T^\mu_{\phantom{\mu}\nu}$ necessary to describe the finer structure present in the universe -- clumping of dark matter, for instance -- once cosmologists try to probe it at smaller scales and higher resolution. These perturbations will also produce inhomogeneities in the metric, so that now
\begin{align}
\label{4DFlatFLRW_Perturbed}
g_{\mu\nu}[\eta,\vec{x}] = a[\eta]^2 \left( \eta_{\mu\nu} + h_{\mu\nu}[\eta,\vec{x}] \right) .
\end{align}
It is possible to perform a scalar-vector-tensor decomposition of both the matter $\delta T^{\alpha}_{\phantom{\alpha}\beta}$ and metric $h_{\alpha\beta}$ fluctuations, such that at linear order in these fields, eq. \eqref{EinsteinEquatin} would yield separate partial differential equations (PDEs) for each of the perturbations transforming differently under the rotation group SO$_3$, an isometry group of the background 4D FLRW geometry in eq. \eqref{4DFlatFLRW}.\footnote{For a detailed and pedagogical treatment of cosmological perturbation theory, see for example, Chapter 5 of Weinberg's cosmology textbook \cite{Weinberg:2008zzc}.}

This paper is specifically about understanding the causal structure of the transverse-traceless (TT) metric perturbations
\begin{align*}
h_{\mu\nu} \dd x^\mu \dd x^\nu = D_{ij} \dd x^i \dd x^j ,
\end{align*}
with $\delta^{ij} D_{ij} = \delta^{ij} \partial_i D_{jk} = 0$. The $D_{ij}$, describing gravitational radiation capable of carrying energy-momentum to infinity, is governed by
\begin{align}
\label{4DFLRW_ScalarPDE}
\overline{\Box} D_{ij} = 16 \pi \GN \Pi_{ij}^\text{(T)} ,
\end{align}
with $\Pi_{ij}^\text{(T)}$ denoting the part of $\delta T^i_{\phantom{i}j}$ that is subject to the TT constraints, $\delta^{ij} \Pi_{ij}^\text{(T)} = \delta^{ij} \partial_i \Pi_{jk}^\text{(T)} = 0$.\footnote{These TT requirements provides 4 equations, reducing the 6 components of the symmetric $D_{ij}$ and of $\Pi_{ij}^\text{(T)}$ down to 2 independent ones.} Now, the $\overline{\Box}$ in \eqref{4DFLRW_ScalarPDE} is the scalar one, i.e.,
\begin{align}
\overline{\Box} D_{ij} = \frac{1}{\sqrt{|\gb|}} \partial_\mu \left( \sqrt{|\gb|} \gb^{\mu\nu} \partial_\nu D_{ij} \right),
\end{align}
where $\sqrt{|\gb|}=a^4$, so that the contribution to the deviation from eq. \eqref{4DFlatFLRW} due to the presence of $\Pi_{ij}^\text{(T)}$ is given by
\begin{align}
\label{4DMetricPerturbationFromSource}
D_{ij}[\eta,\vec{x}]
= 16 \pi \GN \int \dd \eta' \int_{\mathbb{R}^3} \dd^3 \vec{x}' a[\eta']^4 G_4^+[\eta,\vec{x};\eta',\vec{x}'] \Pi^\text{(T)}_{ij}[\eta',\vec{x}'] ,
\end{align}
with the retarded $G_4^+$ obeying
\begin{align}
\overline{\Box}_{\eta,\vec{x}} G_4^+[\eta,\vec{x};\eta',\vec{x}'] 
= \overline{\Box}_{\eta',\vec{x}'} G_4^+[\eta,\vec{x};\eta',\vec{x}']
= \frac{\delta[\eta-\eta']\delta^{(3)}[\vec{x}-\vec{x}']}{a[\eta]^2 a[\eta']^2} .
\end{align}
We see that the study of the causal structure of TT gravitational waves (GWs) propagating in our universe is the same as that of a minimally coupled massless scalar field. By causal structure, we are referring to the fact that, in a curved spacetime, particles that are otherwise massless in 4D Minkowski spacetime no longer travel strictly on the null cone -- they travel both on and within it. (The part of the field traveling inside the light cone is usually called the ``tail".) While this is known in principle \cite{Hadamard}, a thorough understanding of the causal structure of waves propagating in physically important geometries is lacking, particularly in black hole spacetimes. The black hole case is not only a rich problem, because caustics can form from the focusing of null geodesics -- indicating the causal structure of the Green's functions in Kerr spacetime will likely be highly non-trivial -- but understanding it has become fundamental to a successful prediction of GW signatures from Extreme-Mass-Ratio-Inspiral systems.\footnote{See \cite{Yang:2013shb} through \cite{Ori} for a sample of recent work related to Green's functions in Kerr/Schwarzschild spacetimes.}

In cosmology, photons and gravitons are assumed to follow null geodesics, and this is usually justified via JWKB arguments. If the universe is perfectly homogeneous and isotropic, this actually is an exact statement for electromagnetic fields because of the conformal symmetry enjoyed by the 4D Maxwell action. However, as we will see explicitly below, while light requires the inclusion of inhomogeneities in eq. \eqref{4DFlatFLRW_Perturbed} to develop tails \cite{Chu:2011ip}, GWs already do so in the 4D spatially flat FLRW background geometry of eq. \eqref{4DFlatFLRW} -- except during radiation domination. Moreover, current and future generation GW detectors have the potential to listen in on sources at cosmological distances. One may wonder if the tail part of such GW signals can thus acquire additional properties during their propagation, as a result of their interaction with the geometry of the expanding universe, such that they can in turn be used to tell us something about the history of cosmic evolution. The estimates made later on in the paper inform us that, unfortunately, this is unlikely, i.e., the dominant effect of cosmic evolution on $D_{ij}$ is that its waves propagate at unit speed and redshift as $1/a[\eta]$. Nonetheless, we remark that the computation of GW signals from astrophysical sources, such as in-spiraling compact binary systems, are usually performed with asymptotically Minkowskian boundary conditions. A treatment of how GWs from astrophysical systems propagate over such cosmological distances would presumably have to involve drawing a correspondence between the stress-energy-momentum $\Pi^{(\text{T})}_{ij}[\eta,\vec{x}]$ in eq. \eqref{4DMetricPerturbationFromSource} with its counterpart in the post-Minkowskian/Newtonian treatment, by demanding that the far field asymptotic predictions of the latter coincides with the near/intermediate zone ones of the former in equations eq. \eqref{4DMetricPerturbationFromSource}.

One of the key technical goals of this paper is to continue the investigation initiated in \cite{Chu:2013xca} and \cite{Chu:2013hra}, to see if embedding the relevant geometry in some appropriate higher dimensional Minkowski spacetime will aid in understanding the causal structure of waves in the former, since the Green's functions encoding causal structure information is known explicitly in the latter in all dimensions greater or equal to 2. In the current cosmological context, we shall find it useful to exploit the spatial translation and rotation symmetries of the 4D FLRW geometry in eq. \eqref{4DFlatFLRW}, as well as a conformal re-scaling of the massless scalar/graviton $D_{ij}$, to first reduce the problem to a 2D one. Following that, we will show how this allows us to cleanly separate the part of $D_{ij}$ that transmits signals on the light cone from its tail part. In particular, the light cone term of its Green's function -- the ``direct" part, as it is usually known in the gravitational physics literature -- is completely determined by the flat 4D Minkowski counterpart modulated by the conformal re-scaling $1/(a[\eta]a[\eta'])$, where $\eta$ and $\eta'$ are the observation and emission time respectively. The tail part of $D_{ij}$ is what carries physical information about how the graviton interacts with the background spacetime.

In this paper we will, for technical convenience, mostly deal with symmetric -- retarded plus advanced -- Green's functions.\footnote{Retarded Green's functions will be labeled with a $^+$ superscript, and is the symmetric counterpart multiplied by $\Theta[\eta-\eta']$; moreover, throughout this paper, $\Theta[\dots]$ is the Heaviside step function and $\delta[\dots]$ is the Dirac delta-function.} Because the retarded/advanced Green's functions are usually normalized so that they can be said to be sourced by some appropriately defined unit strength spacetime point source, the corresponding PDE for the symmetric Green's function $G[y,y']$ is the field observed at $y^\mu \equiv (\eta,\vec{x})$ sourced by a spacetime point source at $y'^\mu \equiv (\eta',\vec{x}')$ of amplitude 2. Specifically, the symmetric Green's function $G_4[y,y']$ equation for $D_{ij}$ obeys
\begin{align}
\label{4DTTGravitonPDE}
\overline{\Box}_y G_4[y,y'] = \overline{\Box}_{y'} G_4[y,y'] = 2 \frac{\delta[\eta-\eta']\delta^{(3)}[\vec{x}-\vec{x}']}{a[\eta]^2 a[\eta']^2} .
\end{align}
We will define \cite{Burko:2002ge},\cite{Haas:2004kw}
\begin{align}
\label{4DConformalRescaling}
G_4[x,x'] \equiv \frac{\widehat{G}_4[x,x']}{a[\eta] a[\eta']} ,
\end{align}
which will then lead us to the following 4D Minkowski wave equation with a time-dependent potential:
\begin{align}
\label{4DMinkowski_Potential}
\left( \partial_\eta^2 - \delta^{ij} \partial_i \partial_j + U[\eta] \right) \widehat{G}_4 
= \left( \partial_{\eta'}^2 - \delta^{ij} \partial_{i'} \partial_{j'} + U[\eta'] \right) \widehat{G}_4 
= 2 \delta[\eta-\eta']\delta^{(3)}[\vec{x}-\vec{x}'] ,
\end{align}
with
\begin{align}
U[\eta] = -\frac{\ddot{a}[\eta]}{a[\eta]} \equiv -a^{-1} \frac{\dd^2 a}{\dd \eta^2} .
\end{align}
For reference, the Ricci scalar $\overline{\mathcal{R}}$ of the 4D spatially flat FLRW universe in eq. \eqref{4DFlatFLRW} is proportional to this potential $U$: $(a^2/6) \overline{\mathcal{R}} = -\ddot{a}/a$.

In section \eqref{Section_4Dto2D}, we will begin by explaining why any Green's function equation of the form in eq. \eqref{4DMinkowski_Potential}, with a space-independent but otherwise arbitrary potential -- not necessarily arising from a cosmological context -- can be reduced to a 2D one. We will then use this 2D $\to$ 4D prescription to cleanly split the ``direct" part of the cosmological Green's function from its tail from the outset. In both 2D and 4D, the relevant wave equations and the light cone boundary conditions obeyed by the tail functions, will be discussed. We will apply what we have learned, to solve for the TT-GW Green's function in a power law cosmology, and use it to estimate the size of the tail effect in a radiation, matter and cosmological constant dominated universe. In section \eqref{Section_3DScalar}, we offer a different perspective on this 2D wave equation with a potential, by showing how embedding this 2D world in a 4D flat one allows us to see that the 2D Green's function can be sourced by a 3D scalar field. We summarize our findings in section \eqref{Section_Summary}. A significant portion of the analysis in sections \eqref{Section_4Dto2D} and \eqref{Section_3DScalar} is based on appendix \eqref{Section_2DWaveEquation}. There, we discuss the causal structure of the Green's function associated with a flat spacetime 2D wave equation with an arbitrary potential. We will do so directly in 2D and also by embedding the 2D world in a flat 4D one. An infinite Born series solution is derived for the retarded 2D reduced Green's function $\widehat{G}_2^+$ and for its tail function. The light cone boundary conditions obeyed by the first derivatives of this tail function are also worked out. In appendix \eqref{Section_ProofOfBoxTheta} we use position space methods to argue that the symmetric Green's function of the 2D Minkowski wave operator $\partial^2$ is given by $1/2$ everywhere within the light cone of its source and zero outside of the null cone.

\section{4D Causal Structure from 2D}
\label{Section_4Dto2D}

We will begin by explaining why any Green's function $\widehat{G}_4$ in 4D that reflects the spatial translation and rotational symmetries of its background geometry, may be reduced to a 2D problem. Such a high degree of symmetry means both Green's functions ought to depend on the spatial locations of the observers and sources through the Euclidean distances between them. Let us denote $x^1 \equiv x$, $x'^1 \equiv x'$, $\vec{x}_\perp \equiv (0,x^2,x^3)$ and $\vec{x}'_\perp \equiv (0,x'^2,x'^3)$. Next, we proceed to consider
\begin{align}
\label{4Dto2D_Step1}
\widehat{G}_2[\eta,\eta';R] \equiv \int_{\mathbb{R}^2} \dd^2 \vec{x}'_\perp \widehat{G}_4\left[ \eta,\eta'; \sqrt{(x-x')^2 + (\vec{x}_\perp - \vec{x}'_\perp)^2} \right] .
\end{align}
In what follows, it will be useful to work in terms of the symbol $R$. If it occurs within a 2D object, it is $R \equiv |x-x'|$; if it occurs in a 4D object, then it is $R \equiv |\vec{x}-\vec{x}'|$. Similar remarks apply to Synge's world function $\sbar_{y,y'} = (1/2)(y-y')^2$, which is half the square of the geodesic distance between the observer at $y$ and the spacetime point source at $y'$ in flat spacetime. By switching to cylindrical coordinates $r_\perp \equiv |\vec{x}_\perp|$, followed by $\rho \equiv \sqrt{(x-x')^2 + r_\perp^2}$, eq. \eqref{4Dto2D_Step1} then reads
\begin{align}
\widehat{G}_2[\eta,\eta';R] = 2\pi \int_R^\infty \dd \rho \rho \widehat{G}_4[\eta,\eta'; \rho] .
\end{align}
Differentiating both sides with respect to $R$ then tells us $\widehat{G}_4$ does indeed follow from $\widehat{G}_2$:
\begin{align}
\label{4Dfrom2D}
\widehat{G}_4[\eta,\eta'; R] = - \frac{1}{2\pi R} \frac{\partial \widehat{G}_2[\eta,\eta';R]}{\partial R} .
\end{align}
\footnote{This result when applied to Minkowski Green's functions, is why there exists a recursion relation allowing one to construct all even dimensional ones from the 2D case and the odd ones from the 3D result; see \cite{SoodakTiersten} for a pedagogical discussion.}The 2D Green's function, in turn, obeys the 2D analog of eq. \eqref{4DMinkowski_Potential},
\begin{align}
\label{2DMinkowski_Potential}
\left( \partial_\eta^2 - \partial_x^2 + U[\eta] \right) \widehat{G}_2 =
\left( \partial_{\eta'}^2 - \partial_{x'}^2 + U[\eta'] \right) \widehat{G}_2 = 2 \delta[\eta-\eta'] \delta[x-x'] .
\end{align}
To see this, we apply $\mathcal{W}_2 \equiv \partial_\eta^2 - \partial_x^2 + U[\eta]$ to both sides of eq. \eqref{4Dto2D_Step1}. Interchanging the differentiation and integration on the right hand side, and then adding and subtracting the Euclidean Laplacian $\vec{\nabla}_\perp^2 \equiv \partial_2^2 + \partial_3^2$ with respect to $\vec{x}_\perp$, and finally invoking eq. \eqref{4DMinkowski_Potential} obeyed by $\widehat{G}_4$, we have
\begin{align}
\label{4Dto2D_Step2}
\left( \partial_\eta^2 - \partial_x^2 + U[\eta] \right) \widehat{G}_2[\eta,\eta';R] 
= 2 \delta[\eta-\eta']\delta[x-x'] + \int_{\mathbb{R}^2} \dd^2 \vec{x}_\perp \vec{\nabla}_\perp^2 \widehat{G}_4\left[ \eta,\eta'; \sqrt{(x-x')^2 + \vec{x}_\perp^2} \right] .
\end{align}
We recognize the remaining integral on the right hand side of eq. \eqref{4Dto2D_Step2} to be a surface term at spatial infinity. As long as the $\widehat{G}_4$ respects causality, then this surface term must be zero because the source and observer, for fixed times and $x$, $x'$, must lie outside each other's light cone as $|\vec{x}_\perp| \to \infty$. (Similar manipulations hold for the equation with respect to $(\eta',x')$.)

Because of the primary role played by the 2D wave equation \eqref{2DMinkowski_Potential}, we will in appendix \eqref{Section_2DWaveEquation} study its properties for a general potential $U[\xi^{\mu_2} \equiv (\eta,x)]$ -- the reader is invited to take a detour there before returning to the main body of this paper. Here, we will summarize the main results, but specialize to the case where $U$ depends only on the conformal time $\eta$. The solution to eq. \eqref{2DMinkowski_Potential} takes the general pure tail form
\begin{align}
\widehat{G}_2[\eta,x;\eta',x'] = \frac{1}{2} \Theta[\sbar] J[\eta,\eta',R] ,
\end{align}
where $J$ itself obeys the homogeneous 2D wave equation
\begin{align}
\label{2DCosmologyWaveEquation_Homogeneous}
\left( \partial_\eta^2 - \partial_x^2 + U[\eta] \right) J = \left( \partial_{\eta'}^2 - \partial_{x'}^2 + U[\eta'] \right) J = 0 ,
\end{align}
and the light cone boundary condition $J[\sbar=0]=1$. These facts, together with $-(1/R)\partial_R = \partial/\partial \sbar$ and eq. \eqref{4Dfrom2D} immediately imply the 4D reduced Green's function itself takes the form
\begin{align}
\label{4DReducedG_NullConeTailSplit}
\widehat{G}_4 = \overline{G}_4[\sbar] + \frac{\Theta[\sbar]}{4\pi} \frac{\partial J[\eta,\eta';\sbar]}{\partial \sbar} ,
\end{align}
where $\overline{G}_4[\sbar]$ is the 4D minimally coupled massless scalar symmetric Green's function,
\begin{align}
\label{4DFlatMasslessG}
\overline{G}_4[\sbar] = \frac{\delta[\sbar]}{4\pi} .
\end{align} 
Without solving any PDEs, we have managed to isolate the light cone part of $\widehat{G}_4$ from its tail. Note that, because 4D FLRW cosmology $a^2 \eta_{\mu\nu}$ is conformally flat, its light cone is the Minkowskian one $\sbar=0$. Witness too that the 4D tail term $\partial J/\partial \sbar$ in eq. \eqref{4DReducedG_NullConeTailSplit} is entirely determined by the homogeneous solution to the 2D wave equation.
 
Because it may be useful for 4D cosmology, we record here that the analogous homogeneous equations for the 4D graviton tail in terms of $\eta$, $\eta'$ and $\sbar$ treated as independent variables.
\begin{align}
\label{2DWaveEquation_Prime_Homogeneous}
\partial_\eta \left( 2 (\eta-\eta') \partial_{\sbar} J' \right) + \partial_{\sbar} \left( 2 \sbar \partial_{\sbar} J' \right) + (\partial_\eta^2 + U[\eta]) J' &= 0, \\
\partial_{\eta'} \left( 2 (\eta'-\eta) \partial_{\sbar} J' \right) + \partial_{\sbar} \left( 2 \sbar \partial_{\sbar} J' \right) + (\partial_{\eta'}^2 + U[\eta']) J' &= 0. \nonumber
\end{align}
In eq. \eqref{2DWaveEquation_Prime_Homogeneous} and the following paragraphs, $\dot{J} \equiv \partial_\eta J$ and $J' \equiv \partial_{\sbar} J$. As one may expect, evaluating $\sbar = (1/2)((\eta-\eta)'^2 - (x-x')^2 - (\vec{x}_\perp-\vec{x}'_\perp)^2)$, we find that eq. \eqref{2DWaveEquation_Prime_Homogeneous} is simply the homogeneous version of the 4D wave equation for the reduced Green's function (eq. \eqref{4DMinkowski_Potential}),
\begin{align}
\left(\partial_\eta^2 - \delta^{ij} \partial_i \partial_j + U[\eta]\right) J' 
= \left(\partial_{\eta'}^2 - \delta^{ij} \partial_{i'} \partial_{j'} + U[\eta'] \right) J' 
= 0 .
\end{align}
The light cone boundary condition for $J'$ with an arbitrary but space-independent potential $U[\eta]$ is
\begin{align}
\label{2DWaveEquation_Prime_BC}
\left(\frac{\partial J[\eta,\eta';\sbar=0]}{\partial \sbar}\right)_{\eta,\eta'} = -\frac{1}{2(\eta-\eta')} \int_{\eta'}^{\eta} \dd\eta'' U[\eta''] .
\end{align}
In a 4D spatially flat cosmology, $U=-\ddot{a}/a$ and hence eq. \eqref{2DWaveEquation_Prime_BC} yields
\begin{align}
\label{4DCosmologyTailBC}
\left(\frac{\partial J[\eta,\eta';\sbar=0]}{\partial \sbar}\right)_{\eta,\eta'} = \frac{1}{2(\eta-\eta')} \int_{\eta'}^{\eta} \dd\eta'' \frac{\ddot{a}[\eta'']}{a[\eta'']} .
\end{align}
In terms of $\mathcal{W}_2 J \equiv \left( \partial_\eta^2 - \partial_R^2 + U \right) J$, eq. \eqref{2DWaveEquation_Prime_Homogeneous} follows from eq. \eqref{2DWaveEquation_Homogeneous} and
\begin{align}
0 = \partial_{\sbar} \mathcal{W}_2 J = (\mathcal{W}_2 + 2 \partial_{\sbar}) J' .
\end{align}
As for the boundary conditions in eq. \eqref{2DWaveEquation_Prime_BC}, it is convenient to use the light cone coordinates $\xi^\pm \equiv \eta \pm R$ and $\xi'^\pm \equiv \eta'$, so that $\eta = (\xi^+ + \xi^-)/2$ and $\sbar = (1/2)(\xi^+ - \eta')(\xi^- - \eta')$. This in turns implies
\begin{align}
\label{2DWaveEquation_Prime_BC_StepI}
\partial_\pm J &= \frac{1}{2} \dot{J} + \frac{1}{2} (\xi^\mp - \eta') J' .
\end{align}
We now consider the light cone limit $\sbar=0$ by setting $\xi^\pm \to \eta'$. Because $J[\eta,\eta';\sbar=0]=1$ for all times, we must have $\dot{J}[\sbar=0]=0$. Application of equations \eqref{2DLightConeFirstDerivative_J_I} and \eqref{2DLightConeFirstDerivative_J_II} to eq. \eqref{2DWaveEquation_Prime_BC_StepI}, and recognizing $\eta'' = (\xi''^+ + \xi''^-)/2$, returns
\begin{align}
\label{2DWaveEquation_Prime_BC_StepII}
\left(\frac{\xi^\mp + \eta'}{2} - \eta'\right) J'[\eta,\eta';\sbar=0] &= - \frac{1}{2} \int_{\eta'}^{(\xi^\mp + \eta')/2} \dd \eta'' U[\eta''] .
\end{align}
The $\xi^\mp$ are arbitrary at this point, but we would like results expressed solely in terms of $\eta$ and $\eta'$, so we may put $\xi^\pm + \eta' \to 2\eta$ and thereby verify eq. \eqref{2DWaveEquation_Prime_BC}.

What eq. \eqref{4DReducedG_NullConeTailSplit} and eq. \eqref{4DConformalRescaling} inserted to eq. \eqref{4DMetricPerturbationFromSource} teaches us, is that a physical source in any spatially flat FLRW universe produces a null GW front that takes a universal form:
\begin{align}
\label{Dij_Direct}
D_{ij}^{(\gamma)}[\eta,\vec{x}] = 4 \GN
\int_{\mathbb{R}^3} \dd^3 \vec{x}' a[\eta_r]^3 \frac{\Pi^\text{(T)}_{ij} \left[ \eta_r, \vec{x}' \right]}{a[\eta] |\vec{x}-\vec{x}'|} ,
\end{align}
where $\eta_r \equiv \eta-|\vec{x}-\vec{x}'|$ is the retarded time. If the source is isolated, we may define $\vec{x}'=\vec{0}$ to be in its interior. Then in the far field limit -- specifically, whenever $|\vec{x}|$ is much larger than the spatial extent of the source -- we may deduce this null portion of the GW detected at $(\eta,\vec{x})$ is, roughly speaking, the spatial-total of the stress-momentum of the source at retarded time, modulated by the inverse physical distance between observer and source:
\begin{align}
D_{ij}^{(\gamma)}[\eta,\vec{x}] \approx \frac{4 \GN}{a[\eta] |\vec{x}|} 
\int_{\mathbb{R}^3} \dd^3 \vec{x}' a\left[ \eta-|\vec{x}| \right]^3 
\Pi^\text{(T)}_{ij} \left[ \eta-|\vec{x}|, \vec{x}' \right] .
\end{align}
Unlike its direct part, the knowledge of tail $\partial J/\partial \sbar$ requires solving its full wave equation. Hence, it is the part of the GW moving slower than unit speed that actually encodes information regarding how these massless spin-2 particles interact with the background geometry of eq. \eqref{4DFlatFLRW}. 

We now turn to solving the GW tail in a power law cosmology.
\begin{figure}
\begin{center}
\includegraphics[width=4in]{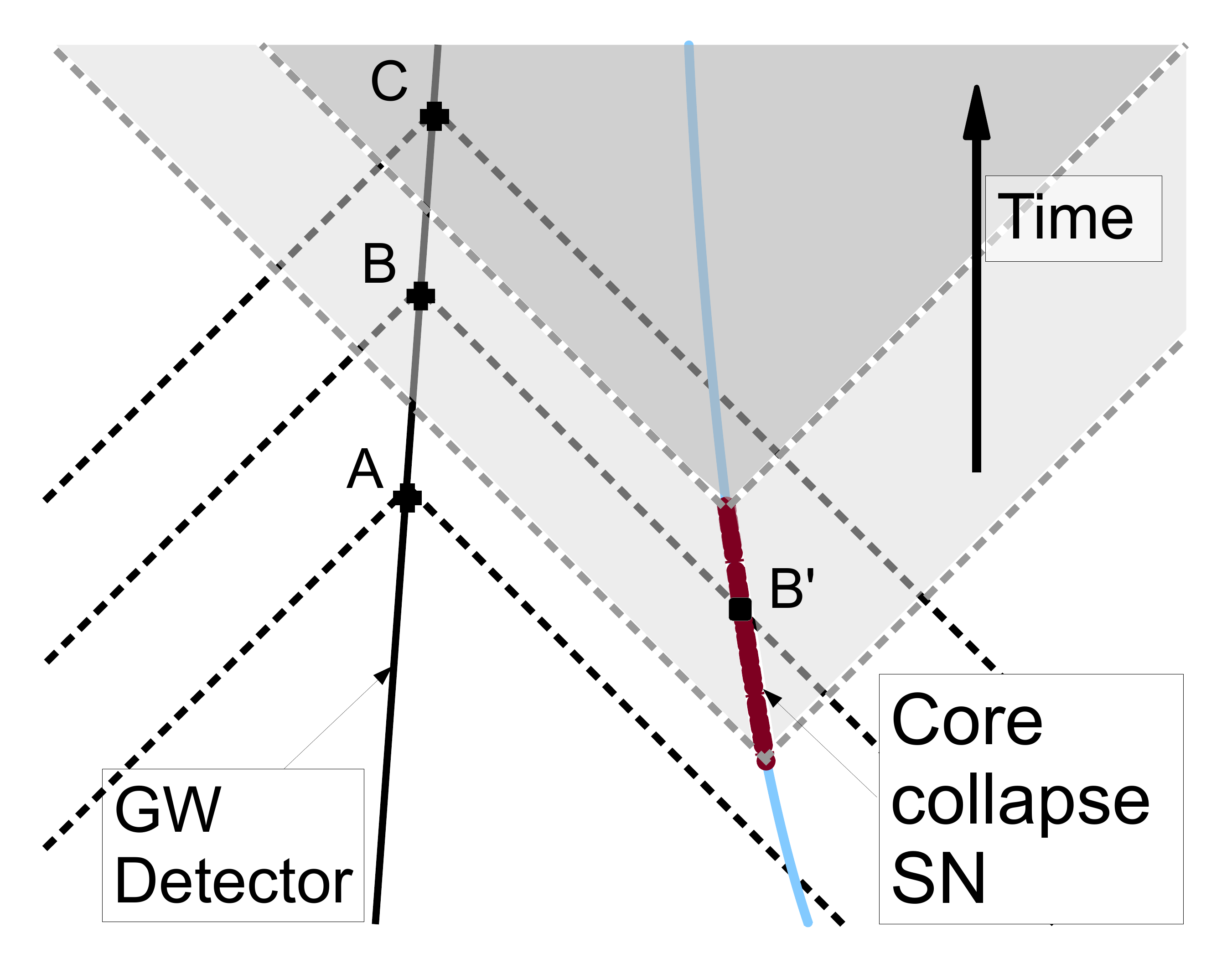}
\caption{{\it Causal structure of TT GWs.} \qquad This spacetime diagram depicts a hypothetical astrophysical process, where a massive star undergoes core collapse and goes supernova (right world line). The dashed-dotted segment of the right world line represents the full duration during which GWs are produced: before that, the collapse has not started; after that, the system has settled down completely. These GWs are detected by a distant detector (left world line). We assume that the background geometry is that of a 4D spatially flat FLRW universe. The black dashed lines emanating from the detector's world line are the past light cones of events $A$, $B$ and $C$. The bottom pair of light gray dashed lines emanating from the right world line is the forward light cone of the beginning of the collapse; the top pair is that of the end of the process. The light gray shaded region of spacetime is filled with GWs propagating both on and inside the null cone. The darker-gray region of spacetime is filled with GW tails only. The detector at $A$ sees no signal. The detector at $B$ sees a ``direct" signal from $B'$ but also a tail from the dashed-dotted segment of the massive star before $B'$. At $C$ the signal received is the accumulation of the GW tails from the entire dashed-dotted segment. As we show in the main text, in a radiation dominated universe, there are in fact no tails, so the detector sees nothing at $C$. In a matter dominated universe, the signal at $C$ is independent of the spatial location but decays with time in an expanding universe. In a de Sitter background, the Green's function tail is a constant; the detector that was operational from $A$ through $C$ would sense a permanent change in $D_{ij}$ that is proportional to $(H/\mpl)^2$ and to the total $\Pi_{ij}^\text{(T)}$ contained in the dashed-dotted segment, i.e., $(H/\mpl)^2 \int \dd^4 x' \sqrt{|\gb[\eta']|} \Pi_{ij}^\text{(T)}[\eta',\vec{x}']$.}
\label{TTGWFigure}
\end{center}
\end{figure}

{\bf Power Law Cosmology} \qquad A scale factor that is a power law in conformal time,
\begin{align}
\label{PowerLaw}
a[\eta] = \left( \frac{\eta}{\eta_0} \right)^p ,
\end{align}
can be used to approximate the major epochs in cosmology -- inflation/dark energy/de Sitter ($p=-1$), radiation ($p=1$) and matter domination ($p=2$). This means $U = -\ddot{a}/a$ becomes
\begin{align}
\label{PowerLawU}
U[\eta] = - \frac{p(p-1)}{\eta^2} ,
\end{align}
and the corresponding 2D wave equation for $J$ in eq. \eqref{2DCosmologyWaveEquation_Homogeneous} can be reduced to an ordinary differential equation. (We would like to attribute this realization to Nariai \cite{Nariai}.\footnote{I also wish to thank Dai De-Chang for bringing \cite{Nariai} to my attention.}) Let $J$ depend on spacetime solely through the combination $s \equiv \sbar/(\eta\eta')$. Then eq. \eqref{2DCosmologyWaveEquation_Homogeneous} becomes
\begin{align}
\label{2DPowerLawCosmology_ODE}
\frac{1}{\eta^2}\left\{ s (s+2) J''[s] + 2 (s+1) J'[s] - (p-1) p J[s] \right\} = 0 .
\end{align}
The regular solution obeying the light cone boundary condition $J[\sbar=s=0] = 1$ reads
\begin{align}
\label{2DSourceForPowerLaw}
J[s] = P_{-p}[s+1] = P_{p-1}[s+1] = \,_2 F_1\left[ 1-p,p;1;-\frac{s}{2} \right], \qquad s \equiv \frac{\sbar}{\eta \eta'} .
\end{align}
We have provided, for the reader's reference, 2 equivalent ways of expressing the Legendre function $P_\nu$, as well as its hypergeometric function $\,_2F_1$ representation; see the $\mu=0$ limit of equations 8.702 and 8.704 of \cite{G&S}. To sum, the symmetric Green's function $\widehat{G}_2$ obeying
\begin{align}
\left(\partial_\eta^2 - \partial_x^2 - \frac{p(p-1)}{\eta^2}\right) \widehat{G}_2[\eta,x;\eta',x'] 
&= \left(\partial_{\eta'}^2 - \partial_{x'}^2 - \frac{p(p-1)}{\eta'^2}\right) \widehat{G}_2[\eta,x;\eta',x'] \nonumber\\
&= 2\delta[\eta-\eta'] \delta[x-x'] ,
\end{align}
is
\begin{align}
\label{2DPowerLaw_GSoln}
\widehat{G}_2[\eta,x;\eta,x'] 
= \frac{1}{2}\Theta[\sbar] P_{-p}\left[ 1 + \frac{\sbar}{\eta \eta'} \right] 
= \frac{1}{2}\Theta[\sbar] \,_2 F_1\left[ 1-p,p;1;-\frac{\sbar}{2 \eta \eta'} \right] ,
\end{align}
with $\sbar \equiv \left\{ (\eta-\eta')^2 - (x-x')^2 \right\}/2$. At this point, putting back the conformal re-scaling (eq. \eqref{4DConformalRescaling}), we have (re-)derived the 4D symmetric Green's function of the minimally coupled massless scalar -- and therefore, also that of the transverse-traceless graviton (obeying eq. \eqref{4DTTGravitonPDE}) -- in a spatially flat FLRW background power law universe (eq. \eqref{PowerLaw}):
\begin{align}
\label{4DTTGraviton_GSolution}
G_4[\eta,\eta';R] 
&= \frac{1}{4\pi a[\eta] a[\eta']} \left( \delta[\sbar] + \Theta[\sbar] \frac{P_{-p}'\left[ 1 + \sbar/(\eta\eta') \right]}{\eta\eta'} \right), \\
&= \frac{1}{4\pi a[\eta] a[\eta']} \left( \delta[\sbar] + \Theta[\sbar] \frac{p(p-1)}{2 \eta \eta'} \,_2 F_1\left[ 2-p,p+1;2;-\frac{\sbar}{2\eta\eta'} \right] \right) , \nonumber
\end{align}
where the prime on the Legendre function denotes derivative with respect to its argument, and here $\sbar \equiv (1/2)((\eta-\eta')^2 - (\vec{x}-\vec{x}')^2)$.\footnote{\label{Footnote_Comparison}We mention here that Caldwell in \cite{Caldwell:1993xw} had claimed to have solved the Green's function of the TT graviton in a FLRW universe with arbitrary spatial curvature. We do not believe his results are correct. When $K=0$ (spatially flat FLRW), the $V/(\eta-\eta')$ in Table I of \cite{Caldwell:1993xw} corresponds to our tail $\partial J/\partial \sbar$. However, on the very first line of Table I, the power law cosmology TT-GW tail obtained there was $\partial J/\partial \sbar = p(p-1)/(2 \eta \eta')$ -- Caldwell's $(\alpha,\eta_i)$ is our $(p,\eta')$ -- and is therefore inconsistent with the solution obtained by \cite{Nariai}, \cite{Haas:2004kw} and in eq. \eqref{4DTTGraviton_GSolution}. Indeed, his spatially flat tail result is really the light cone boundary condition in eq. \eqref{4DCosmologyTailBC_PowerLaw}; compare his (III.6) or (III.7) with eq. \eqref{4DCosmologyTailBC}. One potential source of error is that Caldwell appears to have incorrectly asserted that Synge's world function in FLRW spacetime is given by $\sigma=(1/2)(R^2-(\eta-\eta')^2)$ -- see the statement right before his eq. (III.6) -- which he then used in his eq. (II.7) to calculate the TT GW Green's function. \\ For the reader's reference, we compare the methods used here, \cite{Nariai}, and \cite{Haas:2004kw}, to obtain the minimally coupled massless scalar Green's function in a 4D spatially flat power law cosmology. In eq. (2.13) of \cite{Nariai} and eq. (3.8) of \cite{Haas:2004kw}, a Fourier transform/mode-integral representation involving Bessel functions was found for the Green's function. Nariai \cite{Nariai} replaced the Bessel functions with their asymptotic expansions and proceeded to work out a power series expansion for the Green's function tail in eq. (2.17). Following that, an ansatz for the exact solution was proposed in eq. (3.1), which lead to a hypergeometric (ordinary) differential equation in terms of our $\sbar/(2 \eta\eta')$. The appropriate solution of the two linearly independent ones was then selected by matching it onto the power series in eq. (2.17). In Haas and Poisson \cite{Haas:2004kw}, \cite{G&S} was consulted to directly evaluate the mode integrals in terms of the Appell function $F_4$; see their Green's function tail in eq. (3.10). Hence, their solution for a generic power law cosmology is more complicated that Nariai's \cite{Nariai} and ours (eq. \eqref{4DTTGraviton_GSolution}) in that -- by referring to \cite{G&S}, for instance -- $F_4$ appears to have fewer tabulated properties than $\,_2F_1$. On the other hand, in this paper, we provided a largely self-contained derivation without resorting to the Fourier transform. We showed that the 4D symmetric Green's function is simply related to the first derivative (with respect to $\sbar$) of its 2D cousin. The 2D Green's function, in turn, not only takes a pure tail form, the tail $J$ is fixed uniquely because it satisfies the homogeneous 2D wave equation and is unity on the light cone. To be sure, however, note that our 2D power law solution was gotten by postulating that the tail $J$ was a function solely of $\sbar/(\eta\eta')$, and this was certainly predicated on Nariai's results.} For ease of comparison to Nariai's results in eq. (3.5) of \cite{Nariai}, we also display the solution in terms  $\,_2 F_1$.\footnote{Note that Nariai uses observer time $t$, where $a[t] \propto t^q$ and his background FLRW geometry is $\dd t^2 - a[t]^2 \dd\vec{x}^2$; our $p$ is related to $q$ there via $p = q/(1-q)$; Nariai's $\nu = (3q-1)/(2(1-q))$, $\tau=\eta-\eta'$, $\xi = 2\sbar$ and $z=\sbar/(2\eta\eta')$.} One can check that, since $\,_2 F_1[\alpha,\beta;\gamma;0]=1$, the tail portion of the solution in eq. \eqref{4DTTGraviton_GSolution} satisfies the light cone boundary condition in eq. \eqref{4DCosmologyTailBC}:
\begin{align}
\label{4DCosmologyTailBC_PowerLaw}
\frac{\partial J[\eta,\eta';\sbar=0]}{\partial \sbar} = \frac{p(p-1)}{2 \eta\eta'} .
\end{align}
{\it Radiation domination} \qquad The second line of eq. \eqref{4DTTGraviton_GSolution} makes manifest the fact that there are no GW tails in a radiation dominated ($p=1$) universe. Mathematically this can be understood from $U$ in eq. \eqref{PowerLawU} being zero, and therefore $\widehat{G}_2 = \Theta[\sbar]/2$. One may also recall that, if the radiation were primarily photons, because the Maxwell action is conformally invariant in 4D, the trace of its stress-energy-momentum tensor is zero -- and so is the background Ricci scalar $\overline{\mathcal{R}}$, by Einstein's equations. That $U$ is zero then follows, as we have previously observed that it is proportional to the Ricci scalar.
\begin{align}
\label{4DTTGraviton_Radiation}
G_4^\text{(Radiation)}[\eta,\eta';R] 
&= \frac{\delta[\sbar]}{4\pi a[\eta] a[\eta']}, \qquad \eta,\eta' \in (0,\infty), \qquad a[\eta] = \frac{\eta}{\eta_0} .
\end{align}
Therefore the contribution to the geometry $D_{ij}$ given the TT part of some source $\Pi^\text{(T)}_{ij}$ is completely determined by the universal form in eq. \eqref{Dij_Direct}.

By setting $p=2$ and $p=-1$ in eq. \eqref{4DTTGraviton_GSolution}, one can verify that the reduced 4D Green's function $\widehat{G}_4$ of the TT graviton is in fact the same object in both matter and cosmological constant dominated universe. Their difference in $G_4$ is entirely due to the overall conformal re-scaling $1/(a[\eta] a[\eta'])$. 

{\it Matter domination} \qquad The matter dominated TT graviton Green's function tail is constant in space but goes as $1/(a[\eta]^{\frac{3}{2}} a[\eta']^{\frac{3}{2}})$,
\begin{align}
\label{4DTTGraviton_Matter}
G_4^\text{(Matter)}[\eta,\eta';R] 
&= \frac{1}{4\pi} \left( \frac{\delta[\sbar]}{a[\eta] a[\eta']} + \Theta[\sbar] \frac{\eta_0^4}{(\eta \eta')^3} \right), \qquad \eta,\eta'\in(0,\infty),
\qquad a[\eta] = \left(\frac{\eta}{\eta_0}\right)^2,
\end{align}
and the deviation from eq. \eqref{4DFlatFLRW} due to $\Pi_{ij}^\text{(T)}[\eta',\vec{x}']$ is given by
\begin{align}
D_{ij} = D_{ij}^{(\gamma)} + D_{ij}^\text{(tail)},
\end{align}
where the wavefront that travels at unit speed given by eq. \eqref{Dij_Direct} and the tail part of the GW is
\begin{align}
\label{Cosmology_Matter_Tail}
D_{ij}^\text{(tail)}[\eta,\vec{x}] 
= \frac{4 \GN}{\eta_0^2 a[\eta]^{\frac{3}{2}}} \int_0^{\eta-|\vec{x}-\vec{x}'|-0^+} \dd\eta' a[\eta']^{\frac{5}{2}} \int_{\mathbb{R}^3} \dd^3 \vec{x}' \Pi^\text{(T)}_{ij} \left[ \eta',\vec{x}' \right] .
\end{align}
Suppose the strength of the source peaks at time $\eta_*$, and suppose $\Delta t \sim \int_{\text{peak width}} \dd \eta' a[\eta']$ is the physical duration of the source in the cosmic rest frame. The direct part of the signal in eq. \eqref{Dij_Direct}, in the far field limit, is then roughly bounded by
\begin{align}
\left\vert D_{ij}^{(\gamma)}[\eta,\vec{x}] \right\vert
\lesssim \frac{4 \GN}{a[\eta]|\vec{x}|} 
\left\vert\int_{\mathbb{R}^3} \dd^3 \vec{x}' a[\eta_*]^3 \Pi^\text{(T)}_{ij} \left[ \eta_*,\vec{x}' \right] \right\vert ,
\end{align}
whereas the tail $D_{ij}^\text{(tail)}$ is roughly bounded by 
\begin{align}
\left\vert D_{ij}^\text{(tail)}[\eta,\vec{x}] \right\vert
\lesssim \frac{4 \GN \Delta t}{\eta_0^2 a[\eta]^{\frac{3}{2}} a[\eta_*]^{\frac{3}{2}}} 
\left\vert \int_{\mathbb{R}^3} \dd^3 \vec{x}' a[\eta_*]^3 \Pi^\text{(T)}_{ij} \left[ \eta_*,\vec{x}' \right] \right\vert.
\end{align}
Suppose we normalize the scale factor such that $\eta = \eta_0$ is the observer's time; up to factors of unity, and assuming matter domination throughout cosmic history, $\eta_0$ is also the age of the universe. According to the observer, the ratio of the peak tail amplitude to that of the null cone signal can be estimated as
\begin{align}
\left\vert \frac{D_{ij}^\text{(tail)}[\eta_0,\vec{x}]}{D_{ij}^{(\gamma)}[\eta_0,\vec{x}]} \right\vert
\sim \left(\frac{\Delta t}{\eta_0}\right) \left(\frac{|\vec{x}|}{\eta_0}\right) a[\eta_*]^{-\frac{3}{2}} .
\end{align}
The tail effect, in a matter dominated spatially flat FLRW universe, is suppressed relative to its null cone counterpart, by the ratio of the duration of the source to the age of the universe times the ratio of the observer-source distance (at the observer's time) to the age of the universe. The way to beat this suppression is to have the source of GW reside at a very early epoch of this universe, so that $a[\eta_*] \lesssim ((\Delta t/\eta_0) (|\vec{x}|/\eta_0))^{2/3}$.

{\it de Sitter} \qquad The de Sitter TT graviton Green's function tail is a constant:
\begin{align}
\label{4DTTGraviton_CC}
G_4^\text{($\Lambda$)}[\eta,\eta';R] 
&= \frac{1}{4\pi} \left( \frac{\delta[\sbar]}{a[\eta] a[\eta']} + \frac{\Theta[\sbar]}{\eta_0^2} \right), \qquad \eta,\eta' \in (-\infty,0),
\qquad a[\eta] = \frac{\eta_0}{\eta}.
\end{align}
(This is consistent with the de Sitter minimally coupled massless scalar result in \cite{Chu:2013xca},\cite{Chu:2013hra}.) Therefore the metric fluctuation $D_{ij} =  D_{ij}^{(\gamma)} + D_{ij}^\text{(tail)}$ engendered by the TT part of the source $\Pi^\text{(T)}_{ij}[\eta',\vec{x}']$ has a null cone portion given by eq. \eqref{Dij_Direct} and a tail part that reads
\begin{align}
\label{Cosmology_dS_Tail}
D_{ij}^\text{(tail)}[\eta,\vec{x}] 
= \frac{4 \GN}{\eta_0^2}
\int_{-\infty}^{\eta-|\vec{x}-\vec{x}'|-0^+} \dd\eta' \int_{\mathbb{R}^3} \dd^3 \vec{x}' \sqrt{|\gb[\eta']|} \Pi^\text{(T)}_{ij} \left[\eta',\vec{x}'\right] . 
\end{align}
Assuming $\Pi^\text{(T)}_{ij} \left[\eta',\vec{x}'\right]$ describes a source that radiates GWs over a finite duration, any detector that was present before, during, and after the GW train has passed the observer's location will find that $D_{ij}$ does not decay back to zero but instead suffers a permanent ``DC" shift $\Delta D_{ij}$ proportional to the spacetime volume integral of $\Pi_{ij}^\text{(T)}$:
\begin{align}
\label{Cosmology_dS_Memory}
\Delta D_{ij} &= \frac{4 \GN}{\eta_0^2} H_{ij}, \\
H_{ij} &\equiv \int_{-\infty}^{0^-} \dd\eta' \int_{\mathbb{R}^3} \dd^3 \vec{x}' \sqrt{|\gb[\eta']|} \Pi^\text{(T)}_{ij} \left[\eta',\vec{x}'\right] . 
\end{align}
\footnote{For foundational work on the gravitational memory effect in asymptotically flat spacetimes, see \cite{ZeldovichPolnarev} and \cite{Christodoulou:1991cr}; for more recent theoretical investigations, \cite{Favata:2010zu,Bieri:2013ada,Tolish:2014oda}; for observational searches, see for instance \cite{Wang:2014zls}. In addition, recent work in \cite{Pasterski:2015tva} and \cite{Strominger:2014pwa} have connected this asymptotically Minkowskian memory effect to the low frequency limit of the Ward-Takahashi identities obeyed by graviton scattering amplitudes due to the Bondi-van der Burg-Metzner-Sachs (BMS) symmetry at null infinity, commonly known as Weinberg's soft graviton theorem. We pose the analogous question here: is the memory effect in eq. \eqref{Cosmology_dS_Memory} related to some symmetries associated with asymptotically de Sitter spacetimes? (Note that de Sitter asymptotics are not as well studied as flat ones -- see \cite{Ashtekar:2014zfa}.)}For de Sitter spacetime, we may identify $\eta_0 = -1/H$ to be the negative reciprocal of the Hubble expansion parameter. This means the size of this memory effect is governed by the square of the ratio of Hubble to that of the Planck mass $(H/\mpl)^2$, with $\GN \sim 1/\mpl^2$; while the magnitude of the components of the constant symmetric matrix $H_{ij}$ is, heuristically speaking, the TT-part of the spacetime-total of the stress-momentum of the source.

Just as for the matter dominated case, let us compare the peak amplitudes of the direct signal to the tail. Again, we assume $\Pi_{ij}^\text{(T)}$ itself peaks at $\eta_*$, and the source duration is $\Delta t \sim \int_{\text{peak width}} \dd \eta' a[\eta']$. Then -- recalling $\sqrt{|\gb[\eta']|} = a[\eta']^4$ -- we may estimate that in the far zone
\begin{align}
\left\vert D_{ij}^{(\gamma)}[\eta,\vec{x}] \right\vert
&\lesssim \frac{4 \GN}{a[\eta] |\vec{x}|}
\left\vert \int_{\mathbb{R}^3} \dd^3 \vec{x}' a[\eta_*]^3 \Pi^\text{(T)}_{ij} \left[ \eta_*,\vec{x}' \right] \right\vert \\
\left\vert D_{ij}^\text{(tail)}[\eta,\vec{x}] \right\vert
&\lesssim \frac{4 \GN \Delta t}{\eta_0^2}
\left\vert \int_{\mathbb{R}^3} \dd^3 \vec{x}' a[\eta_*]^3 \Pi^\text{(T)}_{ij} \left[ \eta_*,\vec{x}' \right] \right\vert .
\end{align}
The ratio of the tail amplitude to that of the null cone signal is roughly, $(H \cdot \Delta t) (H \cdot a[\eta]|\vec{x}|)$, where $a|\vec{x}|$ is the co-moving source-observer distance evaluated at the observer's time.

We illustrate what we have found regarding the causal structure of FLRW TT gravitons in Fig. \eqref{TTGWFigure}. We reiterate that the estimates performed here provide strong evidence that, for the most part, TT GWs in our universe propagate on the null cone. The predominant effect of cosmic evolution on them is the redshift $1/a[\eta]$ in eq. \eqref{Dij_Direct}.

\section{3D Scalar Perspective: 2D Embedded in 4D}
\label{Section_3DScalar}

We will now take a different perspective on the 2D wave equation of eq. \eqref{2DMinkowski_Potential}, by demonstrating that it is possible to view the 2D world as embedded in some 4D Minkowski spacetime, such that the $\widehat{G}_2$ is sourced by an appropriate 2D time dependent plane source $J[\eta,\eta';r_\perp^2 \equiv \vec{x}_\perp^2]$:
\begin{align}
\label{2DFrom4DSource}
\widehat{G}_2[\eta,\eta';R] = \int_{\mathbb{R}^2} \dd^2 \vec{x}_\perp \overline{G}_4[\bar{\sigma}] J[\eta,\eta';\vec{x}_\perp^2] , \qquad
\bar{\sigma} \equiv \frac{1}{2}\left( (\eta-\eta')^2 - R^2 - \vec{x}_\perp^2 \right) , 
\end{align}
where we have let the spatial dimension of the 2D world pierce the 2D $\vec{x}_\perp$-plane at its origin $\vec{x}_\perp = \vec{0}_\perp$, and $\overline{G}_4$ (from eq. \eqref{4DFlatMasslessG}) itself obeys
\begin{align}
\partial^2_y \overline{G}_4[\sbar] = \partial^2_{y'} \overline{G}_4[\sbar] = 2 \delta^{(4)}[y-y'] .
\end{align}
As it turns out, $J$ can be viewed as a cylindrically symmetric scalar field living in 3D, obeying
\begin{align}
\label{2DScalarGreensFunction_From4DSource_SourcePDE}
\left( \partial_\eta^2 + \vec{\nabla}_\perp^2 + U[\eta] \right) J + (\eta-\eta')\vec{\nabla}_\perp \ln [r_\perp^2] \cdot \vec{\nabla}_\perp \partial_\eta J = 0, \\
\left( \partial_{\eta'}^2 + \vec{\nabla}_\perp^2 + U[\eta'] \right) J + (\eta'-\eta)\vec{\nabla}_\perp \ln [r_\perp^2] \cdot \vec{\nabla}_\perp \partial_{\eta'} J = 0 . \nonumber
\end{align}
and the boundary condition
\begin{align}
\label{2Dfrom4D_J_BC}
J[\eta,\eta';\vec{x}_\perp = \vec{0}_\perp] = 1, \qquad\qquad \forall \eta,\eta'.
\end{align}
We discuss in appendix \eqref{Section_2DWaveEquation} the more general case where the $U$ depends on both $\eta$ and the first spatial coordinate $x$; we refer the reader to it for the derivation of equations \eqref{2DScalarGreensFunction_From4DSource_SourcePDE} and \eqref{2Dfrom4D_J_BC}.
\begin{figure}
\begin{center}
\includegraphics[width=4in]{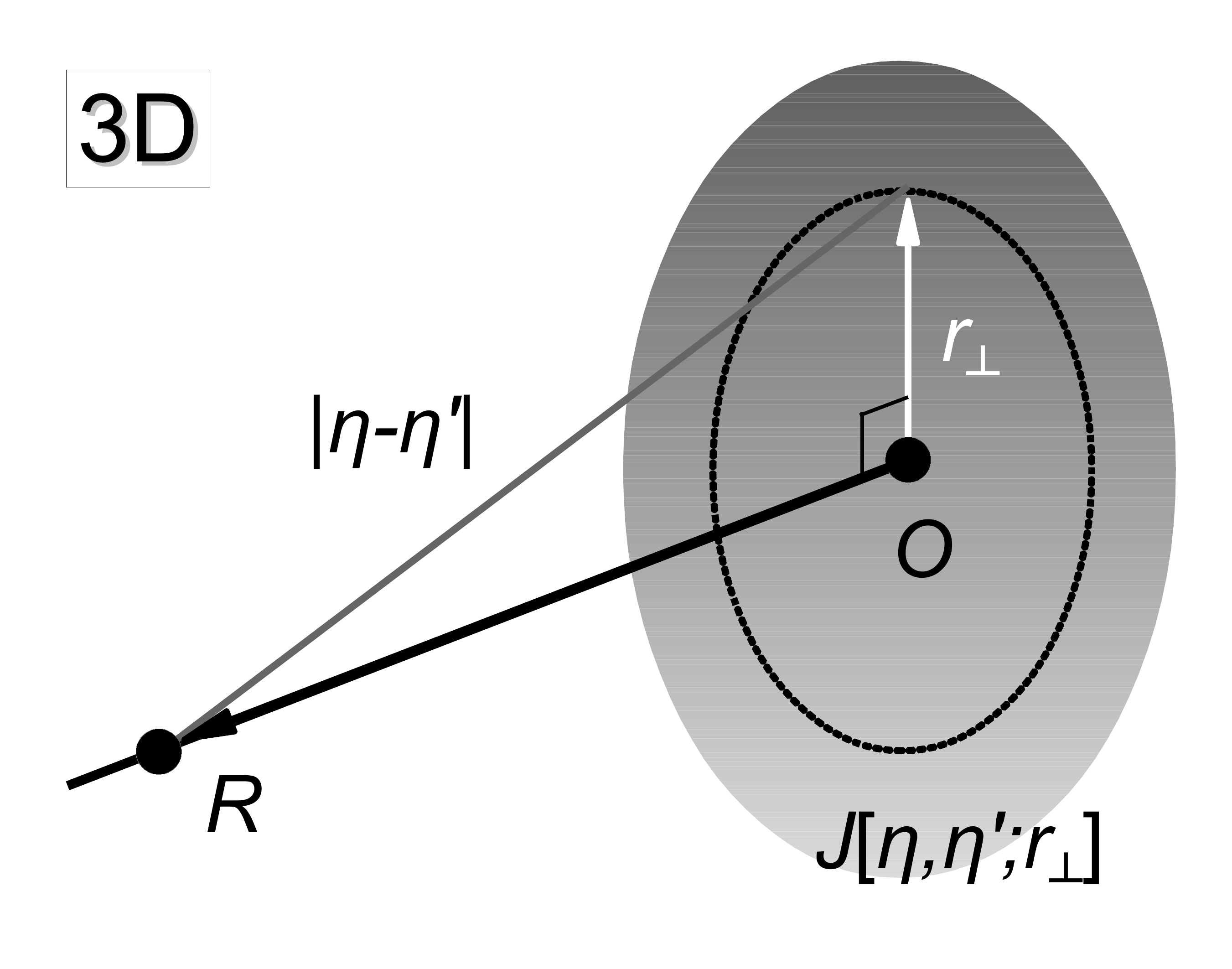}
\caption{{\it 1D space embedded in 3D space.} \qquad This figure illustrates the causal structure encapsulated in the integral representation of $\widehat{G}_2$ in eq. \eqref{2DFrom4DSource}. The large shaded oval is to be viewed as a portion of the infinite 2D plane source $J[\eta,\eta';r_\perp]$, which comes into existence for only an instant at $\eta'$; whereas the observer's time is $\eta$, i.e., the elapsed time between observation and emission is $\eta-\eta'$. The 1D space of the 2D world whose scalar waves are described by $\widehat{G}_2$, pierces the 2D plane source $J$ orthogonally at its origin (denoted by $O$). The 2D observer is located $R$ away from $O$ along the 1D line. (In 2D $R\equiv|x-x'|$, where $x$ and $x'$ are, respectively, the spatial locations of the observer and the source of the Green's function. In 4D $R\equiv|\vec{x}-\vec{x}'|$, with similar meanings for $\vec{x}$ and $\vec{x}'$.) Because the ambient 4D spacetime is Minkowskian (the $\overline{G}_4$ in eq. \eqref{2DFrom4DSource}), for a fixed $R$ and elapsed time $|\eta-\eta'|$, no signal from any part of the 2D plane source $J$ can reach the observer whenever $|\eta-\eta'| < R$. Suppose instead $|\eta-\eta'| \geq R$, then because massless scalars in 4D Minkowski travel strictly on the light cone, the observed signal -- from this 4D perspective -- receives contributions only from the (dotted) circle on $J$ of infinitesimal thickness $\dd r_\perp$ and radius defined by $|\eta-\eta'|^2 = R^2 + r_\perp^2$. But from the 2D point of view $|\eta-\eta'| > R$ is simply the statement that the observed signal is the tail of $\widehat{G}_2$; from eq. \eqref{2DFrom4DSource_J} we also see the 2D signal has no $\delta$-function impulse at $|\eta-\eta'|=R$. Finally, since it is the integrated signal that is being observed, there is no need to allow for the source $J$ to have an azimuthal dependence.}
\label{2DTailFigure}
\end{center}
\end{figure} 
Notice from the prescription in eq. \eqref{2DFrom4DSource}, that since $\overline{G}_4$ itself is Poincar\'{e} invariant, we have attributed to the source $J$ all the time-translation symmetry breaking effects encoded in $\widehat{G}_2$, due to the presence of the potential $U[\eta]$. The representation in eq. \eqref{2DFrom4DSource} also illuminates the causal structure of a signal due to a spacetime point source at $(\eta',x')$ in the 2D world, in terms of that of the collective 4D signal due to the 2D plane source $J[\eta,\eta';\vec{x}_\perp^2]$ -- we explain it in some detail through Fig.\eqref{2DTailFigure}.\footnote{It it worth highlighting that the embedding perspective here should not be taken too literally because the source $J[\eta,\eta';r_\perp]$, although turned on only for an instant, somehow needs to take into account the time of observation $\eta$ to properly fool the 2D observer she is detecting the field of a spacetime point source at $x'$. This ``acausality" of the source never arises when solving Green's functions within a given geometry; for example, in $\Box G[X,X'] = \delta^{(d)}[X-X']/\sqrt[4]{|g[X]g[X']|}$ the source location $X'$ is completely independent of the observer's $X$.} Furthermore, by using eq. \eqref{4DFlatMasslessG} and switching to cylindrical coordinates in eq. \eqref{2DFrom4DSource}, we may demonstrate explicitly that $\widehat{G}_2[\eta,\eta';R]$ is pure tail (see eq. \eqref{G2_PureTailIntegral}):
\begin{align}
\label{2DFrom4DSource_J}
\widehat{G}_2[\eta,\eta';R] &= \frac{\Theta[\sbar]}{2} J[\eta,\eta';r_\perp^2 = 2\sbar] .
\end{align}
Utilizing $-(1/R)\partial_R = \partial_{\sbar}$ and equations \eqref{4Dfrom2D} and \eqref{4DFlatMasslessG} on eq. \eqref{2DFrom4DSource_J}, we again recover the null-cone versus tail split in eq. \eqref{4DReducedG_NullConeTailSplit}. Moreover, when the relationship $r_\perp \leftrightarrow 2 \sbar$ is kept in mind, a direct calculation will show that equations \eqref{2DScalarGreensFunction_From4DSource_SourcePDE} and \eqref{2Dfrom4D_J_BC} is in fact just a re-phrasing of the 2D homogeneous equation obeyed by $J[\eta,\eta';\sbar]$, namely
\begin{align}
\label{2DHomoEqn}
& J^{(2,0,0)}[\eta,\eta';R] - J^{(0,0,2)}[\eta,\eta';R] + U[\eta] J[\eta,\eta';R]  \\
&= J^{(0,2,0)}[\eta,\eta';R] - J^{(0,0,2)}[\eta,\eta';R] + U[\eta'] J[\eta,\eta';R] 
= 0  \nonumber
\end{align}
and the light cone boundary condition $J[\eta,\eta';r_\perp^2=2\sbar=0]=1$ -- except here $R \equiv \sqrt{(\eta-\eta')^2-r_\perp^2}$. 

Is there some general expansion scheme that would allow us to solve eq. \eqref{2DScalarGreensFunction_From4DSource_SourcePDE} using, for e.g., a separation-of-variables technique, while incorporating the boundary condition eq. \eqref{2Dfrom4D_J_BC}? At the moment we merely pose the question, but will show how the power law universe results of the past section follows readily from equations \eqref{2DScalarGreensFunction_From4DSource_SourcePDE} and \eqref{2Dfrom4D_J_BC}.

{\bf Power law cosmology} \qquad From the 3D scalar point of view, when the scale factor of the universe is a power law in conformal time $a[\eta] = (\eta/\eta_0)^p$, we use the ansatz $J[s \equiv r_\perp^2/(2 \eta \eta')]$, and will find $J$ obeys, via eq. \eqref{2DScalarGreensFunction_From4DSource_SourcePDE},
\begin{align}
\frac{4 \eta'^2 s^2}{r_\perp^4} \left(s (s+2) J''[s]+2 (s+1) J'[s]-(p-1) p J[s]\right) = 0.
\end{align} 
This is, of course, equivalent to eq. \eqref{2DPowerLawCosmology_ODE}. Requiring $J[r_\perp=0]=1$, eq. \eqref{2Dfrom4D_J_BC}, leads to the solution
\begin{align}
J[\eta,\eta',r_\perp] = P_{-p}[s+1] = P_{p-1}[s+1] = \,_2 F_1\left[ 1-p,p;1;-\frac{s}{2} \right], \qquad s \equiv \frac{r_\perp^2}{2\eta \eta'} .
\end{align}
We recover the results of the previous section once we invoke eq. \eqref{2DFrom4DSource_J}.

\section{Summary, Discussion, and A 5D Question}
\label{Section_Summary}

In this paper we have shown that, after the (known) conformal re-scaling of the TT graviton Green's function in a 4D spatially flat FLRW universe, an understanding of its causal structure can be achieved by first reducing the problem to a 2D one. It turns out that the reduced 4D Green's function's ``direct" part is equal to its Minkowski cousin, for any background cosmic history $a[\eta]$. It is really the tail part of the TT graviton that has physical information about how these massless spin 2 particles interact with the background FLRW geometry, since it is here that the wave equation needs to be solved in detail before the tail can be known explicitly. What we have uncovered is the 4D graviton tail function can be thought of as the derivative of the tail of the associated 2D Green's function with respect to $\sbar$, Synge's world function in Minkowski spacetime. It can also be viewed as the radial derivative of a cylindrically symmetric scalar field residing in 3D. This 3D perspective comes from embedding the 2D world in a 4D Minkowski spacetime. The 2D Green's function is sourced by some instantaneous (at $\eta'$) 2D plane source $J[\eta,\eta';|\vec{x}_\perp|]$, which intersects the 1D space of the 2D world orthogonally at $x'$. This $J$ is a 3D scalar because it depends on both time and the 2D space parametrized by $\vec{x}_\perp$. Moreover, while in the 2D context the tail function obeys boundary condition on the light cone -- it is unity there -- the analogous boundary condition for the 3D scalar $J$ translates to a spatial one, where $J[\eta,\eta';|\vec{x}_\perp|=0]=1$. 

We have built on earlier work by Nariai \cite{Nariai}, and used our 2D and 3D pictures to recover the exact TT graviton Green's function in a power law cosmology. We found there are no TT GW tails in a radiation dominated universe. In a matter dominated cosmology, the TT GW tail is constant in space but decays with increasing time (in conformal coordinates). We estimated, in this case, that the ratio of the peak amplitude of the tail part of the signal generated by an isolated source, to that of the direct part, to be proportional to the ratio of the duration of the source to the age of the universe and also to the observer-source distance (at the observer's time) to the same. In de Sitter spacetime, the tail of the TT graviton Green's function is a constant, and this leads to an inside-the-null-cone memory effect proportional to $(H/\mpl)^2 \sim \GN H^2$, and to the TT-portion of the spacetime-total stress-momentum of the source. On the other hand, in 4D Minkowski spacetime, the massless scalar and the spin$-2$ graviton travel strictly on the light cone, i.e., they have no tails. As far as causal structure is concerned, therefore, the de Sitter memory effect uncovered here has no analog to its counterpart(s) in 4D asymptotically flat spacetimes.

As part of our explorations, we have studied in some detail -- see appendix \eqref{Section_2DWaveEquation} -- the properties of the 2D Minkowski wave equation with an arbitrary potential. This includes not only the pure tail nature of the corresponding 2D Green's function; the light cone boundary conditions obeyed by the tail function $J$ and its first derivatives; an infinite Born series solution for both the retarded version of the 2D Green's function and for $J$; as well as how the 2D tail $J$ can also be viewed as a 3D scalar field by embedding the 2D world in 4D Minkowski. As a by product, we have found that there are at least 3 other ways of obtaining the 2D Minkowski massive scalar Green's function, apart from its Fourier integral representation: by reducing the associated homogeneous wave equation into an ordinary differential equation; by switching to the 3D scalar picture, where the tail obeys a Laplace-Helmholtz equation; and by evaluating the tail's infinite Born series expansion term-by-term.

Dimension reduction, together with embedding, has allowed us to write down a PDE solely for the massless scalar/TT-graviton tail, before it is solved for explicitly. This is to be contrasted with what is usually done in cosmology, where a Fourier expansion of Einstein's equations linearized about the background geometry of eq. \eqref{4DFlatFLRW} is performed from the outset. Fourier space is crucial, of course, for applications such as predicting the power spectra of metric fluctuations that could be traced back to the inflationary epoch of the early universe. But to obtain information regarding the causal structure of TT-GWs, one would have to inverse Fourier transform the mode expansion of the Green's function, and except for special cases where the mode functions are known explicitly -- and simple enough for the integrals to be performed -- momentum space actually appears to hide both the general features of the light cone versus tail separation and the detailed solution of the tail itself. In fact our recognition here, that the Green's function of the massless scalar field in 4D spatially flat FLRW can be cleanly split into its direct and tail pieces, can also be found in past literature -- see for example, eq. (3.3) of \cite{Haas:2004kw} (which was a follow up of \cite{Burko:2002ge}). However, in \cite{Haas:2004kw}, where the case of a power law cosmology $a[\eta] = (\eta/\eta_0)^\alpha$ was solved, the tail part of the Green's function was still gotten via the Fourier transform of the whole Green's function (equations (3.4), (3.6) and (3.8)). We see that working in position spacetime as far as possible, and utilizing the dimension reduction/embedding perspective, have let us arrive more readily at a simpler form of the solution consistent with that found by Nariai \cite{Nariai}.\footnote{It could be that some integral transform or expansion scheme is useful for the solution of the tail part of the cosmological scalar Green's function, but this we advocate, should be done directly for the PDE obeyed by the 2D tail itself, not the entire 4D Green's function. Also, since the tail portion of the cosmological Green's function, when reduced to the 2D context, is unity on the light cone, we suggest here that any useful method should be able to incorporate this boundary condition in a transparent manner.} (See footnote \eqref{Footnote_Comparison} for the more subtle details.)

We actually started out this 4D spatially flat FLRW investigation motivated by the fact that our 4D universe can be embedded in 5D Minkowski. Is there a ``line mass/charge" in 5D piercing the 4D FLRW world at $x'$, such that the 4D observer is fooled into thinking that she is detecting the field generated by a spacetime point source at $x'$, i.e., that of a Green's function $G_4[x,x']$? This is still unanswered. We hope, however, that the insights we have gained in this paper would help us further this goal.

\section{Acknowledgments}

I had exchanges with numerous people on the GW tail effect in cosmology. A partial list includes: Thorsten Battefeld, Lasha Berezhiani, Chiara Caprini, Dai De-Chang, \'{E}anna Flanagan, Abraham Harte, Mudit Jain, Justin Khoury, Eric Poisson, and Vitaly Vanchurin. Much of the analytic work was carried out with {\sf Mathematica} \cite{Mathematica}. This work began while I was a postdoctoral fellow at the University of Pennsylvania, supported by NSF PHY-1145525 and funds from the University of Pennsylvania. Much of the analytic work performed in this paper was carried out using {\sf Mathematica} \cite{Mathematica}.

\appendix

\section{2D Minkowski Wave Equation With An Arbitrary Potential}
\label{Section_2DWaveEquation}

In this section, we wish to examine the dynamics of a scalar field obeying a 2D Minkowski wave equation with a potential $U$ which is both time- and space-dependent. The defining equation for the symmetric (retarded plus advanced) Green's function ($\widehat{G}_2[\xi,\xi'] = \widehat{G}_2[\xi',\xi]$) reads
\begin{align}
\label{2DWaveEquation}
\left(\partial_\xi^2 + U[\xi]\right) \widehat{G}_2[\xi,\xi'] 
= \left(\partial_{\xi'}^2 + U[\xi']\right) \widehat{G}_2[\xi,\xi']
= 2\delta^{(2)}[\xi-\xi'] . 
\end{align}
$\xi^{\mu_2} \equiv (\eta,x)$ and $\xi'^{\mu_2} \equiv (\eta',x')$ denote coordinates in $(1+1)$D; the subscript $2$ in $\mu_2$ indicate the spacetime indices run from $0$ to $1$, so $\partial^2 = \partial^{\mu_2} \partial_{\mu_2} = \partial_\eta^2 - \partial_x^2$ and $\partial^{\mu'_2} \partial_{\mu'_2} = \partial_{\eta'}^2 - \partial_{x'}^2$, for instance. The key observation is that the solution to eq. \eqref{2DWaveEquation} takes the generic pure tail form 
\begin{align}
\label{2DWaveEquation_GeneralSolution}
\widehat{G}_2[\xi,\xi'] = \frac{1}{2} \Theta[\sbar] J[\xi,\xi'] , \qquad 
\sbar \equiv \frac{1}{2}(\xi-\xi')^2 ,
\end{align}
where $J$ obeys the homogeneous wave equation
\begin{align}
\label{2DWaveEquation_Homogeneous}
\left(\partial_\xi^2 + U[\xi]\right)J[\xi,\xi'] 
= \left(\partial_{\xi'}^2 + U[\xi']\right) J[\xi,\xi']
= 0, 
\end{align}
and the boundary condition that it is unity on the light cone
\begin{align}
\label{2DWaveEquation_BC}
J[\sbar=0] = 1.
\end{align}
Here, $\sbar$ is half the square of the geodesic distance between $\xi$ and $\xi'$ in 2D Minkowski. Because we are dealing with the symmetric Green's function, $J[\xi,\xi'] = J[\xi',\xi]$. The step function in eq. \eqref{2DWaveEquation_GeneralSolution} tells us scalar waves in 2D obeying eq. \eqref{2DWaveEquation} travel strictly inside the cone of its physical sources. This is to be contrasted with the 4D case, where there is an additional term proportional to $\delta[\sbar]$, telling us scalar waves produced by a spacetime point source also receive contributions from a impulsive shock wavefront traveling exactly with unit speed.

We will now proceed to derive equations \eqref{2DWaveEquation_GeneralSolution} through \eqref{2DWaveEquation_BC} by a direct calculation. Insert eq. \eqref{2DWaveEquation_GeneralSolution} into eq. \eqref{2DWaveEquation}. On the left-hand-side, finds
\begin{align}
\label{2DWaveEquation_Step1}
(\partial^2+U)\widehat{G}_2 &= \frac{1}{2}\left(\partial^2 \Theta \cdot J + 2 \delta[\sbar] \partial_\mu \sbar \partial^\mu J + \Theta \cdot (\partial^2 + U) J \right) .
\end{align}
We will argue in appendix \eqref{Section_ProofOfBoxTheta} that, in Cartesian coordinates,
\begin{align}
\label{2DWaveEquation_BoxThetaIdentity}
\partial^2_\xi \Theta[\sbar] = \partial^2_{\xi'} \Theta[\sbar] = 4 \delta^{(2)}[\xi-\xi'] .
\end{align}
Thus, the first term from the left of eq. \eqref{2DWaveEquation_Step1} returns the desired $\delta$-functions in the right hand side of eq. \eqref{2DWaveEquation} if $J=1$ when $\xi=\xi'$. Moreover, for eq. \eqref{2DWaveEquation} to hold, the coefficients of the $\delta$- and $\Theta$-function in eq. \eqref{2DWaveEquation_Step1} needs to vanish. Noting $\partial^\mu \sbar = (\xi-\xi')^\mu$, the $\delta$-function coefficient informs us $J$ must be constant on the light cone:
\begin{align}
(\xi-\xi')^\mu \partial_\mu J[\sbar=0] = 0 .
\end{align}
But since $J$ is unity at the apex of the light cone $\xi=\xi'$, that means it must in fact be unity everywhere $\sbar=0$, i.e., we have eq. \eqref{2DWaveEquation_BC}. For the $\Theta$ function term to vanish everywhere inside the light cone, we see $J$ must obey the homogeneous equation in eq. \eqref{2DWaveEquation_Homogeneous}.

{\it Massive Scalar in 2D Minkowski} \qquad As an application, let us derive the massive scalar Green's function in 2D Minkowski,
\begin{align}
\label{2DMassiveScalarG_PDE}
\left(\partial^2+U\right)\overline{G}_2[\xi-\xi'] = 2 \delta^{(2)}[\xi-\xi'], \qquad U = m^2.
\end{align}
We will assume, because of the highly symmetric nature of the problem, that $J = J[m\sqrt{2\sbar}]$ depends on $\xi$ and $\xi'$ solely through $\sbar$ alone. The homogeneous equation \eqref{2DWaveEquation_Homogeneous} then reads
\begin{align}
\label{2DMassiveG_PDEtoODE}
m^2 \left( J''[\chi] + \frac{J'[\chi]}{\chi} + J[\chi] \right) = 0, \qquad \chi \equiv m \sqrt{2\sbar} .
\end{align}
The solution that obeys the light cone boundary condition $J[\chi,\sbar \to 0]=1$ is the Bessel function $J_0[m\sqrt{(\xi-\xi')^2}]$. We have, therefore, the solution to the massive scalar symmetric Green's function:
\begin{align}
\label{2DMassiveG_Solution}
\overline{G}_2[\xi-\xi'] = \frac{1}{2} \Theta[\sbar] J_0\left[ m \sqrt{2 \sbar}\right] .
\end{align}
This result can be cross-checked by performing the Fourier integral
\begin{align}
\Theta[\eta-\eta']\overline{G}_2[\xi-\xi'] = \int_{\text{ret}} \frac{\dd^2 k}{(2\pi)^2} \frac{e^{-ik_\mu (\xi-\xi')^\mu}}{-k^2 + m^2} ,
\end{align}
where the retarded contour needs to be chosen for the $k_0$-integral.

{\bf 2D Embedded in 4D} \qquad Let us imagine that our 2D world is embedded in 4D Minkowski, and ask if there is some source in 4D that could in fact yield $\widehat{G}_2$. Employing the minimally coupled massless scalar 4D Green's function $\overline{G}_4$ in eq. \eqref{4DFlatMasslessG}, we will now argue that
\begin{align}
\label{2DScalarGreensFunction_From4DSource}
\widehat{G}_2[\xi,\xi'] &= \int_{\mathbb{R}^2} \dd^2 \vec{x}_\perp \overline{G}_4[\bar{\sigma}] J[\xi,\xi';\vec{x}_\perp^2] , \qquad\qquad
\bar{\sigma} \equiv \frac{1}{2}\left( (\xi-\xi')^2 - \vec{x}_\perp^2 \right) ,
\end{align}
where the source $J$ obeys the equation(s)
\begin{align}
\label{2DScalarGreensFunction_From4DSourcePDE}
\left( \partial_{\mu_2} \partial^{\mu_2} + \vec{\nabla}_\perp^2 + U[\xi] \right) J[\xi,\xi';\vec{x}_\perp^2]
+ \vec{\nabla}_\perp \ln [\vec{x}_\perp^2] \cdot \vec{\nabla}_\perp \left((\xi-\xi')^{\mu_2}\partial_{\mu_2} J[\xi,\xi';\vec{x}_\perp^2]\right) = 0, \\
\left( \partial_{\mu_2'} \partial^{\mu_2'} + \vec{\nabla}_\perp^2 + U[\xi'] \right) J[\xi,\xi';\vec{x}_\perp^2]
+ \vec{\nabla}_\perp \ln [\vec{x}_\perp^2] \cdot \vec{\nabla}_\perp \left((\xi'-\xi)^{\mu_2}\partial_{\mu_2'} J[\xi,\xi';\vec{x}_\perp^2]\right) = 0, \nonumber
\end{align}
and the boundary condition at the spatial origin of the 2D-$\vec{x}_\perp$ plane
\begin{align}
\label{2DScalarGreensFunction_From4DSourcePDE_BC}
J[\xi,\xi';\vec{x}_\perp = \vec{0}_\perp] = 1 \qquad\qquad \forall \xi,\xi'.
\end{align}
Notice, if this prescription is valid, we have attributed to the source $J$ all the 2D Poincar\'{e} symmetry breaking due to the presence of the potential $U$; for otherwise, $\overline{G}_4$ is itself Poincar\'{e} invariant. Also, for some fixed elapsed time $\eta-\eta'$ and fixed source spatial location $x'$, the signal at $x$ is going to come from integrating over the circle on the $\vec{x}_\perp$-plane defined by $|\eta-\eta'|^2=(x-x')^2+\vec{x}_\perp^2$, or equivalently $\vec{x}_\perp^2 = 2\sbar$; and since it is integrated over anyway, there is no need to allow for the source $J$ to have an azimuthal dependence. Note that eq. \eqref{2DScalarGreensFunction_From4DSource} provides a second means to deduce that $\widehat{G}_2$ is pure tail. Recalling eq. \eqref{4DFlatMasslessG},
\begin{align}
\label{G2_PureTailIntegral}
\widehat{G}_2[\xi,\xi'] 
&= \int_0^\infty \frac{\dd r_\perp^2}{2} \delta\left[ (\xi-\xi')^2-r_\perp^2 \right] J[\xi,\xi';r_\perp^2] \\
&= \frac{\Theta[\sbar]}{2} J\left[ \xi,\xi';r_\perp = \sqrt{2\sbar} \right] . \nonumber
\end{align}
\footnote{The integral in eq. \eqref{G2_PureTailIntegral} is a special case of $\int_{z_1}^{z_2} \dd z \delta[z-a]f[z]=f[a]\Theta[z_2-a]\Theta[a-z_1]$; the top hat $\Theta[z_2-a]\Theta[a-z_1]$ is the constraint that, the integral returns $f[a]$ if $a$ lies within the interval $a \in [z_1,z_2]$ and zero otherwise.}We begin the derivation of  equations \eqref{2DScalarGreensFunction_From4DSourcePDE} and \eqref{2DScalarGreensFunction_From4DSourcePDE_BC} by highlighting that $\overline{G}_4[\sbar]$ itself depends on spacetime solely through the Minkowski world function $\sbar$. Now, apply the 2D wave operator on both sides of eq. \eqref{2DScalarGreensFunction_From4DSource}, interchange the order of integration and differentiation, and add and subtract $\vec{\nabla}_\perp^2$.
\begin{align}
\left( \partial_{\mu_2} \partial^{\mu_2} + U\right) \widehat{G}_2[\xi,\xi'] 
= \int_{\mathbb{R}^2} \dd^2 \vec{x}_\perp \Big\{
& J \left( \partial_{\mu_2} \partial^{\mu_2} - \vec{\nabla}_\perp^2 \right) \overline{G}_4 \\	
&+ \left( \overline{G}'_4 \cdot 2 \partial_{\mu_2} \sbar \partial^{\mu_2} J + \overline{G}_4 \cdot ( \partial_{\mu_2} \partial^{\mu_2} + \vec{\nabla}_\perp^2 + U) J \right) 
\Big\} . \nonumber
\end{align}
We have integrated-by-parts one of the $\vec{\nabla}^2_\perp$ onto the $J$, and also used the fact that $\overline{G}_4$ depends on spacetime solely through $\sbar$. In fact, let us observe that
\begin{align}
-\vec{x}_\perp \cdot \vec{\nabla}_\perp \overline{G}_4 = (-)^2 \vec{x}^2_\perp \cdot \overline{G}'_4 ,
\end{align}
and thus the prime on the $\overline{G}_4$ can be expressed in terms of $\vec{\nabla}_\perp$ and then integrated-by-parts:
\begin{align}
\int \dd^2 \vec{x}_\perp \overline{G}'_4 \cdot 2 \partial_{\mu_2} \sbar \partial^{\mu_2} J
&= 2 \int \dd^2 \vec{x}_\perp \overline{G}_4 \vec{\nabla}_\perp \cdot \left( \frac{\vec{x}_\perp}{\vec{x}^2_\perp} \partial_{\mu_2} \bar{\sigma} \partial^{\mu_2} J \right) .
\end{align}
Because $\vec{\nabla}_\perp \ln \vec{x}^2_\perp = 2 \vec{x}_\perp/\vec{x}^2_\perp$,
\begin{align}
2 \vec{\nabla}_\perp \cdot \left( \frac{\vec{x}_\perp}{\vec{x}^2_\perp} \partial_{\mu_2} \bar{\sigma} \partial^{\mu_2} J \right)
&= \vec{\nabla}_\perp \ln [\vec{x}_\perp^2] \partial_{\mu_2} \bar{\sigma} \partial^{\mu_2} \cdot \vec{\nabla}_\perp J
	+ \vec{\nabla}^2_\perp \ln [\vec{x}_\perp^2] \partial_{\mu_2} \sbar \partial^{\mu_2} J .
\end{align}
The Green's function equation for the 2D Laplacian
\begin{align}
\vec{\nabla}^2_\perp \ln[\vec{x}^2_\perp] = 4\pi \delta^{(2)}[\vec{x}_\perp] ,
\end{align}
together with the 4D Minkowski wave equation $\partial^2 \overline{G}_4 = 2\delta^{(4)}$, then allow us to gather
{\allowdisplaybreaks\begin{align}
\label{2DScalarGreensFunction_From4DSourcePDE_LastStep}
\left(\partial_{\mu_2} \partial^{\mu_2} + U\right) \widehat{G}_2[\xi,\xi'] 
&= 2 J[\xi=\xi';\vec{0}_\perp] \delta^{(2)}[\xi-\xi']
	+ 4\pi \overline{G}_4[\bar{\sigma}] \cdot (\xi-\xi')^{\mu_2} \partial_{\mu_2} J[\xi,\xi';\vec{0}_\perp] \\
&\qquad\qquad
	+ \int_{\mathbb{R}^2} \dd^2 \vec{x}_\perp \overline{G}_4 \left\{
	\left( \partial_{\mu_2} \partial^{\mu_2} + \vec{\nabla}_\perp^2  + U \right) J 
	+ \vec{\nabla}_\perp \ln [\vec{x}_\perp^2] \partial_{\mu_2} \bar{\sigma} \cdot \vec{\nabla}_\perp \partial^{\mu_2} J
	\right\} . \nonumber
\end{align}}
To obtain $2\delta^{(2)}[\xi-\xi']$ on the right hand side, we demand that eq. \eqref{2DScalarGreensFunction_From4DSourcePDE} be satisfied to have the second line of eq. \eqref{2DScalarGreensFunction_From4DSourcePDE_LastStep} vanish, and the simultaneous boundary conditions $J[\xi=\xi';\vec{0}_\perp]=1$ and $(\xi-\xi')^{\mu_2} \partial_{\mu_2} J[\xi,\xi';\vec{0}_\perp]=0$ for all $\xi$, $\xi'$. But these latter two conditions amount to demanding $J[\xi,\xi';\vec{0}_\perp]=1$ for any $\xi$, $\xi'$, i.e., eq. \eqref{2DScalarGreensFunction_From4DSourcePDE_BC}. 

{\it 2D Minkowski Massive Scalar} \qquad We now show how this 3D picture offers an alternate perspective on eq. \eqref{2DMassiveScalarG_PDE}, the 2D massive scalar flat spacetime Green's function $\overline{G}_2$, which through eq. \eqref{4Dfrom2D}, would yield the latter's 4D counterpart. We have $U=m^2$, and will further make an inspired guess, based on the correspondence between $2\sbar$ and $r_\perp^2$ (recall eq. \eqref{G2_PureTailIntegral}), that $J$ is actually time independent -- since in 2D, Poincar\'{e} invariance indicates $\overline{G}_2$ depends on spacetime solely through $\sbar$. (In this specific context, there is a relationship between rotational O$_2$ invariance in the 3D picture and Poincar\'{e} symmetry in 2D.) The problem of massive waves in 2D Minkowski is now translated into, via eq. \eqref{2DScalarGreensFunction_From4DSourcePDE}, a static Laplace-Helmholtz type equation in 2 spatial dimensions:
\begin{align}
\left( \vec{\nabla}_\perp^2 + m^2 \right)J[r_\perp] = 0.
\end{align}
(This reduces to the same equation in eq. \eqref{2DMassiveG_PDEtoODE}.) The solution obeying the boundary condition in eq. \eqref{2Dfrom4D_J_BC}, $J[r_\perp=0]=1$ is the Bessel function 
\begin{align}
J[r_\perp] = J_0[mr_\perp] .
\end{align}
Inserting this into eq. \eqref{G2_PureTailIntegral} yields the result in eq. \eqref{2DMassiveG_Solution}.

{\bf First derivatives on the light cone} \qquad  Let us employ light cone coordinates $\xi^\pm \equiv \xi^0 \pm \xi^1$, so that the 2D flat metric reads $\dd \xi^+ \dd \xi^-$ and $\partial^2 = 4 \partial_+ \partial_-$. For some scalar field $\psi$ obeying the homogeneous 2D wave equation $(\partial^2 + U) \psi = 0$, suppose its value is known on the light cone based at some fixed location $\xi'$, i.e., $\psi[(\xi-\xi')^2=0]$ is known; this also means the first derivatives $\partial_+ \psi[\xi^+,\xi^-=\xi'^-]$ and $\partial_- \psi[\xi^+=\xi'^+,\xi^-]$ along the light cone can be derived. (In fact, by this assumption, all higher derivatives $\partial_+^n \psi[\xi^+,\xi^-=\xi'^-]$ and $\partial_-^n \psi[\xi^+=\xi'^+,\xi^-]$, $n \geq 1$, can be computed.) The other first derivatives of $\psi$ evaluated on the light cone of $\xi'$ is given by the following integrals:
\begin{align}
\label{2DLightConeFirstDerivative_I}
\partial_+ \psi[\xi^+=\xi'^+,\xi^-] 
&= \partial_+ \psi[\xi=\xi'] 
- \frac{1}{4} \int_{\xi'^-}^{\xi^-} \dd \xi''^- U[\xi^+=\xi'^+, \xi''^-] \psi[\xi^+=\xi'^+,\xi''^-], \\
\partial_- \psi[\xi^+,\xi^-=\xi'^-] 
\label{2DLightConeFirstDerivative_II}
&= \partial_- \psi[\xi=\xi'] 
- \frac{1}{4} \int_{\xi'^+}^{\xi^+} \dd \xi''^+ U[\xi''^+, \xi^-=\xi'^-] \psi[\xi''^+,\xi^-=\xi'^-] 
\end{align}
To see this, we first observe that the relevant boundary conditions at the light cone's apex $\xi=\xi'$ are satisfied. Next, we take a derivative with respect to $\partial_-$ on both sides of equations \eqref{2DLightConeFirstDerivative_I} and $\partial_+$ on \eqref{2DLightConeFirstDerivative_II}. This recovers the wave equation $\partial_+ \partial_- \psi = -(1/4) U \psi$ evaluated on the light cone.

Equations \eqref{2DLightConeFirstDerivative_I} and \eqref{2DLightConeFirstDerivative_II}, when applied to the tail function of $\widehat{G}_2$ respecting $J[\sbar=0]=1$ (eq. \eqref{2DWaveEquation_BC}), hands us
\begin{align}
\label{2DLightConeFirstDerivative_J_I}
\partial_+ J[\xi^+ = \xi'^+, \xi^-] 
&= - \frac{1}{4} \int_{\xi'^-}^{\xi^-} \dd \xi''^- U[\xi^+=\xi'^+, \xi''^-], \\
\label{2DLightConeFirstDerivative_J_II}
\partial_- J[\xi^+,\xi^-=\xi'^-] 
&= - \frac{1}{4} \int_{\xi'^+}^{\xi^+} \dd \xi''^+ U[\xi''^+, \xi^-=\xi'^-] .
\end{align}
The first derivatives of $J$ at the null cone's apex $\xi=\xi'$ is zero because $J$ is a constant on the entire light cone.

{\bf Born Series Solution for $\widehat{G}_2^+$ and $J$} \qquad We now turn to an infinite Born series solution for the retarded Green's function $\widehat{G}_2^+$ in 2D, in terms of its counterpart $\overline{G}^+_2$ (which obeys $\partial^2 \overline{G}_2^+ = \delta^{(2)}$) and the potential $U$. This will also yield a corresponding series for $J$ itself. We first define the operator $\mathcal{Q}$ acting on a bi-scalar $S_{\xi,\xi'}$ as
\begin{align}
\label{OperatorQ}
\left(\mathcal{Q} S\right)[\xi,\xi'] \equiv \int_{\mathbb{R}^{(1,1)}} \dd^2 \xi'' \overline{G}_2[\xi-\xi''] U[\xi''] S[\xi'',\xi'] , \qquad \text{(Cartesian)}.
\end{align}
Then the infinite Born series is
\begin{align}
\label{BornSeries_G2}
\widehat{G}_2^+[\xi,\xi'] = \sum_{\ell=0}^{\infty} (-)^\ell \left(\mathcal{Q}^\ell \overline{G}_2^+\right)[\xi,\xi'] .
\end{align}
The zeroth term is defined to be $\left(\mathcal{Q}^{\ell=0} \overline{G}_2^+\right)[\xi,\xi'] \equiv \overline{G}_2^+[\xi-\xi']$. For $\ell \geq 1$, and written in light cone coordinates $\xi^\pm \equiv \eta \pm x$ and $\dd^2 \xi \to (1/2) \dd \xi^+ \dd \xi^-$,
\begin{align}
\label{QSeries}
& \left(\mathcal{Q}^\ell \overline{G}_2^+\right)[\xi,\xi'] 
= \frac{1}{2^\ell}\left( \prod_{s=1}^{\ell} \int_{\mathbb{R}^{(1,1)}} \dd \xi_s^+ \dd \xi_s^- \right) \\
&\times
\overline{G}_2^+[\xi-\xi_\ell] U[\xi_\ell] \overline{G}_2^+[\xi_\ell-\xi_{\ell-1}] U[\ell_{\ell-1}] \overline{G}_2^+[\xi_{\ell-1}-\xi_{\ell-2}] 
\dots \overline{G}_2^+[\xi_2-\xi_1] U[\xi_1] \overline{G}_2^+[\xi_1-\xi'] . \nonumber
\end{align}
These $\ell$-nested spacetime integrals can be further broken down into,
\begin{align}
\label{ISeries}
\left(\mathcal{Q}^\ell \overline{G}_2^+\right)[\xi,\xi'] = \frac{\overline{G}_2^+[\xi,\xi']}{4^\ell} \mathcal{I}_\ell[\xi,\xi'] ,
\end{align}
where $\mathcal{I}_0[\xi,\xi'] \equiv 1$; using the shorthand $\int_{\xi'}^{\xi} \dd^2 \xi''^\pm \equiv \int_{\xi'^+}^{\xi^+} \dd \xi''^+ \int_{\xi'^-}^{\xi^-} \dd \xi''^-$, and for $\ell \geq 1$,
\begin{align}
\label{ISeries_NestedIntegrals}
\mathcal{I}_\ell[\xi,\xi']
&\equiv \int_{\xi'}^{\xi} \dd^2 \xi_\ell^\pm U[\xi_\ell] \int_{\xi'}^{\xi_\ell} \dd^2 \xi_{\ell-1}^\pm U[\xi_{\ell-1}] \dots \int_{\xi'}^{\xi_3} \dd^2 \xi_2^\pm U[\xi_2] \int_{\xi'}^{\xi_2} \dd^2 \xi_1^\pm U[\xi_1] , \\
\label{ISeries_Recursion}
&= \int_{\xi'}^{\xi} \dd^2 \xi_\ell^\pm U[\xi_\ell] \mathcal{I}_{\ell-1}[\xi_\ell,\xi'] . 
\end{align}
The Born series for $J$ itself, whenever $\xi$ lies within the future light cone of $\xi'$ -- which we will denote as $\xi > \xi'$ -- is
\begin{align}
\label{BornSeries_J}
J[\xi > \xi'] = \sum_{\ell=0}^{\infty} \frac{(-)^\ell}{4^\ell} \mathcal{I}_\ell[\xi,\xi'] .
\end{align}
For the purposes of solving the 2D Green's function $\widehat{G}_2$ we do not need $J$ outside the light cone. Therefore, once $J[\xi,\xi']$ is known for $\xi$ lying within the future light cone of $\xi'$, by symmetry this means we have obtained $J[\xi',\xi]$ for $\xi'$ inside the past light cone of $\xi$; by swapping the labels $\xi \leftrightarrow \xi'$ this means we also know $J[\xi,\xi']$ for $\xi$ lying in the past light cone of $\xi'$.

{\it Derivation} \qquad Let us now justify equations \eqref{BornSeries_G2} through \eqref{BornSeries_J}. We begin by applying $\partial_\xi^2$ and $\partial_{\xi'}^2$ to eq. \eqref{QSeries}. For $\ell=0$, the result is just the $\delta$-function $\delta^{(2)}[\xi-\xi']$. For $\ell=1$, the result is
\begin{align}
\partial^2 \mathcal{Q} \overline{G}_2^+ = U \overline{G}_2^+ .
\end{align}
For $\ell > 1$, the leftmost or rightmost integral will collapse, returning
\begin{align}
\partial^2 \mathcal{Q}^\ell \overline{G}_2^+ = U \mathcal{Q}^{\ell-1} \overline{G}_2^+ .
\end{align}
A direct calculation would therefore tell us
\begin{align}
\partial^2 \widehat{G}^+_2 
= \delta^{(2)}[\xi-\xi'] - U \sum_{\ell=0}^{+\infty} (-)^\ell \left(\mathcal{Q}^\ell \overline{G}_2^+\right) .
\end{align}
which is $(\partial^2 + U)\widehat{G}_2 = \delta$, the retarded version of eq. \eqref{2DWaveEquation}.\footnote{As an aside, we remark here that the derivation up to this point is not specific to 2D. This cosmological 2D reduced Green's function problem is, in fact, not the only curved spacetime example that admits a solution as an infinite Born series in $\overline{G}$, its corresponding Minkowski Green's function, because of the ``flat $\Box$"-plus-potential form of the wave operator. The other instance we are aware of, is that of the minimally coupled massless scalar and photon retarded Green's functions in black hole (BH) geometries put in a Kerr-Schild form \cite{KSPaper}. There, for both Schwarzschild and Kerr spacetimes, the $\ell$th term in the summation goes as (BH mass)$^\ell$ times $\ell$-nested spacetime integrals that are independent of the BH mass but involve for the Kerr case the BH's angular momentum.} Next, we will show that using the retarded flat 2D massless scalar Green's function in this construction implies the solution for $\widehat{G}_2^+$ also obeys the retarded boundary condition. Now, for any object $U$, the integral
\begin{align}
\label{Q1}
\left(\mathcal{Q} \overline{G}_2^+\right)[\xi,\xi'] 
= \frac{1}{2}\int_{\mathbb{R}^{(1,1)}} \dd^2 \xi''^\pm \frac{\Theta_+[\xi,\xi'']}{2} U[\xi''] \frac{\Theta_+[\xi'',\xi']}{2}
\end{align}
is zero if $\xi$ lies outside the future light cone of $\xi'$, due to the presence of the 2 retarded Green's functions in the integrand. We have also used the result
\begin{align}
\overline{G}_2^+[\xi-\xi'] = \frac{1}{2}\Theta[\eta-\eta']\Theta[\sbar_{\xi,\xi'}] \equiv \frac{1}{2}\Theta_+[\xi,\xi'] .
\end{align}
When $\xi > \xi'$, the integral in eq. \eqref{Q1} is really that of $U[\xi'']/4$ over the rectangular region on the $\xi''$-2D plane bounded by the past light cone of $\xi$ and the future light cone of $\xi'$, i.e., $(1/4)\mathcal{I}_1[\xi,\xi']$. It is zero otherwise. What we have argued, therefore, is that 
\begin{align}
\label{Q1_Prime}
\left(\mathcal{Q} \overline{G}_2^+\right)[\xi,\xi'] = \frac{\overline{G}_2^+[\xi-\xi']}{4} \mathcal{I}_1[\xi,\xi'] .
\end{align}
Let us suppose that eq. \eqref{ISeries} holds for some $\ell$. Then 
\begin{align}
\label{Qell+1}
\left(\mathcal{Q}^{\ell+1} \overline{G}_2^+\right)[\xi,\xi'] = \frac{1}{4^\ell \cdot 2} 
\int_{\xi'}^{\xi} \dd^2 \xi_{\ell+1}^\pm \overline{G}_2^+[\xi-\xi_{\ell+1}] U[\xi_{\ell+1}] \mathcal{I}_\ell[\xi_{\ell+1},\xi'] \overline{G}_2^+[\xi_{\ell+1}-\xi'] .
\end{align}
The causality arguments employed to analyze eq. \eqref{Q1} tell us, because of the two retarded Green's functions, the $\mathcal{Q}^{\ell+1} \overline{G}_2^+$ must again be proportional to $\overline{G}_2^+[\xi-\xi']$. Assuming $\xi > \xi'$ the integral in eq. \eqref{Qell+1} becomes that of $U[\xi_{\ell+1}] \mathcal{I}_\ell[\xi_{\ell+1},\xi']/4$ over the rectangular region bounded by the past and future null cones of $\xi$ and $\xi'$ respectively; eq. \eqref{Qell+1} is thus
\begin{align}
\left(\mathcal{Q}^{\ell+1} \overline{G}_2^+\right)[\xi,\xi'] = \frac{\overline{G}_2^+[\xi-\xi']}{4^{\ell+1}} 
\int_{\xi'}^{\xi} \dd^2 \xi_{\ell+1}^\pm U[\xi_{\ell+1}] \mathcal{I}_\ell[\xi_{\ell+1},\xi']  .
\end{align}
Recalling the integral recursion relation between $\mathcal{I}_{\ell}$ and $\mathcal{I}_{\ell-1}$ in eq. \eqref{ISeries_Recursion} then proves eq. \eqref{ISeries} for arbitrary $\ell \geq 1$.

Because we have shown that every term in the Born series of $\widehat{G}_2^+$ in eq. \eqref{BornSeries_G2} is proportional to $\overline{G}_2^+$, we may in fact define $J$ to be the latter's coefficient, namely
\begin{align}
\widehat{G}_2^+[\xi,\xi'] = \overline{G}_2^+[\xi-\xi'] J[\xi,\xi'] = \overline{G}_2^+[\xi-\xi'] \sum_{\ell=0}^{\infty} \frac{(-)^\ell}{4^\ell} \mathcal{I}_\ell[\xi,\xi'] .
\end{align}
(The first equality is, of course, consistent with eq. \eqref{2DWaveEquation_GeneralSolution}.) The result in eq. \eqref{BornSeries_J} follows once we take $\xi$ to lie within the future light cone of $\xi'$, for there $\overline{G}_2^+[\xi-\xi'] \to (1/2)$.

A different means of arriving at eq. \eqref{BornSeries_J} is to integrate with respect to both $\xi^\pm$ the homogeneous equation $\partial_+ \partial_- J = -(1/4) UJ$ from the light cone to some arbitrary point $\xi$ lying within the future light cone of $\xi'$, taking into account the boundary conditions $J[\sbar=0]=1$ and $\partial_\pm J[\xi^\mp = \xi'^\mp]=0$. The result is the integral equation
\begin{align}
\label{IntegralEquationForJ}
J[\xi > \xi'] = 1 - \frac{1}{4} \int_{\xi'}^{\xi} \dd^2 \xi''^\pm U[\xi''] J[\xi'',\xi'] .
\end{align}
Iterating eq. \eqref{IntegralEquationForJ} infinite times yields eq. \eqref{BornSeries_J}.

{\it 2D Minkowski Massive Scalar Revisited} \qquad For the massive scalar in 2D Minkowski, $U=m^2$ is a constant and every term of eq. \eqref{BornSeries_J} can be evaluated. The $\ell$th term is, in light cone coordinates,
\begin{align}
\frac{m^{2\ell}}{(-4)^\ell} \int_{\xi'}^{\xi} \dd^2 \xi_\ell \int_{\xi'}^{\xi_\ell} \dd^2 \xi_{\ell-1} \dots 
\int_{\xi'}^{\xi_3} \dd^2 \xi_{2} \int_{\xi'}^{\xi_{2}} \dd^2 \xi_{1} .
\end{align}
The rightmost integral gives $(\xi_{2}^+-\xi'^+) (\xi_{2}^--\xi'^-)$. The one after that gives us $(\xi_{3}^+-\xi'^+)^2 (\xi_{3}^--\xi'^-)^2/2^2$, and so on. After all $\ell$ integrals are done we have $(\xi^+-\xi'^+)^\ell (\xi^--\xi'^-)^\ell/(\ell!)^2$. This leads us to
\begin{align*}
J[\xi,\xi'] = \sum_{\ell=0}^{\infty} \frac{m^{2\ell} (\xi^+ - \xi'^+)^{\ell} (\xi^- - \xi'^-)^{\ell}}{(-4)^\ell (\ell!)^2} .
\end{align*}
This is nothing but the infinite series representation of $J_0[m \sqrt{(\xi^+-\xi'^+)(\xi^--\xi'^-)}]$.

\section{Why $\partial^2 \Theta[\sbar] = 4 \delta^{(2)}[\xi-\xi']$.}
\label{Section_ProofOfBoxTheta}

The primary goal in this section is to justify eq. \eqref{2DWaveEquation_BoxThetaIdentity}. Via a direct calculation, with $\partial_\mu \sbar = (\xi-\xi')_\mu$ and $(\xi-\xi')^2 = 2\sbar$,
\begin{align}
\partial^2 \Theta[\sbar] = 2 \partial_{\sbar} \left( \sbar \delta[\sbar] \right) = 0.
\end{align}
Therefore $\Theta[\sbar]$ is the homogeneous solution to the 2D massless scalar wave equation almost everywhere. Let us employ light cone coordinates,
\begin{align}
\xi^\pm \equiv \eta \pm x, \qquad \xi'^\pm \equiv \eta' \pm x' ,
\end{align}
so that $\sbar = (1/2)(\xi^+ - \xi'^+)(\xi^- - \xi'^-)$ and $\partial^2 = 4 \partial_+ \partial_-$. Note that $\Theta[\sbar]$ is a constant inside the light cone, so $\partial^2 \Theta[\sbar]$ can only be non-zero on the light cone, when either $\xi^+ = \xi'^+$ or $\xi^- = \xi'^-$. To argue that it is in fact only non-trivial at the apex of the light cone, i.e., $\xi = \xi'$, we shall first let $\xi^+$ be arbitrary, but integrate the left-hand-side of eq. \eqref{2DWaveEquation_BoxThetaIdentity} about $\xi^- \approx \xi'^-$:
\begin{align}
\label{ProofOfBoxTheta_StepI}
\int_{\xi'^- - 0^+}^{\xi'^- + 0^+} \dd \xi^- 4 \partial_{-} \partial_{+} \Theta[\sbar]
&= \left. \delta[\xi^+ - \xi'^+] 4 \cdot \text{sgn}[\xi^- - \xi'^-] \right\vert_{\xi^- = \xi'^- - 0^+}^{\xi^- = \xi'^- + 0^+} \nonumber\\
&= 8 \delta[\xi^+ - \xi'^+] .
\end{align}
We have used the distributional identity $\delta[ab] = \delta[b]/|a|$. Similarly,
\begin{align}
\label{ProofOfBoxTheta_StepII}
\int_{\xi'^+ - 0^+}^{\xi'^+ + 0^+} \dd \xi^+ 4 \partial_{+} \partial_{-} \Theta[\sbar]
= 8 \delta[\xi^- - \xi'^-] .
\end{align}
Therefore, $\partial^2 \Theta[\sbar]$ is non-zero only when both $\xi^+ \to \xi'^+$ and $\xi^- \to \xi'^-$ simultaneously. This justifies the $\delta^{(2)}[\xi-\xi']$ (written in Cartesian coordinates) on the right hand side of eq. \eqref{2DWaveEquation_BoxThetaIdentity}. The factor of $4$ can now be checked by re-expressing the $\delta$-functions of eq. \eqref{2DWaveEquation_BoxThetaIdentity} in light cone coordinates, $\delta^{(2)}[\xi-\xi'] = 2\delta[\xi^+ - \xi'^+] \delta[\xi^- - \xi'^-]$, and then integrating it with respect to $\dd \xi^+ \dd \xi^-$ over an infinitesimal region around $\xi = \xi'$ to obtain $8$. On the other hand, eq. \eqref{ProofOfBoxTheta_StepI} is already the $\partial^2 \Theta$ integrated with respect to $\dd\xi^-$, so to recover this $8$ one simply has to integrate over $\dd\xi^+$. The same check can be made by integrating both sides of  eq. \eqref{ProofOfBoxTheta_StepII} with respect to $\dd\xi^-$.

\end{document}